\documentclass[iop]{emulateapj}

\usepackage{graphicx}
\usepackage{amsmath}
\usepackage{natbib}
\usepackage{mathptmx}
\usepackage{apjfonts}
\usepackage{times}
\usepackage{multirow}
\usepackage[colorlinks=true,urlcolor=blue,citecolor=blue,linkcolor=blue]{hyperref}
\usepackage{aas_macros}
\usepackage{makecell}

\bibpunct{(}{)}{;}{a}{}{,}

\usepackage[position=t,singlelinecheck=on,font={rm,bf,up},font=large]{subfig}

\makeatletter
\long\def\frontmatter@title@above{
  \vspace*{-17mm}\vspace*{\headheight}
   \hspace{-4.6mm}{\sc The Astrophysical Journal Supplement Series}, 229:10 (14pp), 2017\\
   \vspace*{4mm}\hspace{-1.3mm}{\footnotesize {\sc Preprint typeset using \LaTeX\ style emulateapj}}
  \par\vspace*{-\baselineskip}\vspace{6mm}
  }

\makeatother

\RequirePackage{color}

\shorttitle{High-frequency oscillations in small magnetic elements}
\shortauthors{Jafarzadeh et al.}

\begin{document}

\title{High-frequency Oscillations in Small Magnetic Elements Observed with {\sc Sunrise}/SuFI}

\author{S.~Jafarzadeh\hyperlink{}{\altaffilmark{1}}}
\author{S.~K.~Solanki\hyperlink{}{\altaffilmark{2,3}}}
\author{M.~Stangalini\hyperlink{}{\altaffilmark{4}}}
\author{O.~Steiner\hyperlink{}{\altaffilmark{5,6}}}
\author{R.~H.~Cameron\hyperlink{}{\altaffilmark{2}}}
\author{S.~Danilovic\hyperlink{}{\altaffilmark{2}}}

\affil{\altaffilmark{1}\hspace{0.2em}Institute of Theoretical Astrophysics, University of Oslo, P.O. Box 1029 Blindern, NO-0315 Oslo, Norway; \href{mailto:shahin.jafarzadeh@astro.uio.no}{shahin.jafarzadeh@astro.uio.no}\\
\altaffilmark{2}\hspace{0.2em}Max Planck Institute for Solar System Research, Justus-von-Liebig-Weg 3, D-37077 G\"{o}ttingen, Germany\\
\altaffilmark{3}\hspace{0.2em}School of Space Research, Kyung Hee University, Yongin, Gyeonggi 446-701, Republic of Korea\\
\altaffilmark{4}\hspace{0.2em}INAF-Osservatorio Astronomico di Roma, I-00040 Monte Porzio Catone (RM), Italy\\
\altaffilmark{5}\hspace{0.2em}Kiepenheuer-Institut f\"{u}r Sonnenphysik, Sch\"{o}neckstr. 6, D-79104 Freiburg, Germany\\
\altaffilmark{6}\hspace{0.2em}Istituto Ricerche Solari Locarno, Via Patocchi 57, 6605 Locarno-Monti, Switzerland}

\begin{abstract}

We characterize waves in small magnetic elements and investigate their propagation in the lower solar atmosphere from observations at high spatial and temporal resolution.
We use the wavelet transform to analyze oscillations of both horizontal displacement and intensity in magnetic bright points found in the $300$~nm and the Ca~{\sc ii}~H $396.8$~nm passbands of the filter imager on board the {\sc Sunrise} balloon-borne solar observatory.
Phase differences between the oscillations at the two atmospheric layers corresponding to the two passbands reveal upward propagating waves at high frequencies (up to $30$ mHz). Weak signatures of standing as well as downward propagating waves are also obtained.
Both compressible and incompressible (kink) waves are found in the small-scale magnetic features. The two types of waves have different, though overlapping, period distributions.
Two independent estimates give a height difference of approximately $450\pm100$~km between the two atmospheric layers sampled by the employed spectral bands. This value, together with the determined short travel times of the transverse and longitudinal waves provide us with phase speeds of $29\pm2$~km\,s$^{-1}$ and $31\pm2$~km\,s$^{-1}$, respectively. We speculate that these phase speeds may not reflect the true propagation speeds of the waves. Thus, effects such as the refraction of fast longitudinal waves may contribute to an overestimate of the phase speed.

\vspace{4mm}
\end{abstract}


\section{Introduction}
\label{sec-intro}

Wave phenomena in solar magnetic features have been proposed as a prime means of transferring energy from the solar interior and lower atmosphere to the place where it is needed to maintain the high temperatures of the mid and upper solar atmosphere as well as to power the solar wind~\citep[e.g.,][]{Choudhuri1993b,DePontieu2007,Straus2008,BelloGonzalez2009,Taroyan2009,Vigeesh2009,McIntosh2011,Jafarzadeh2013a}. Due to the fact that acoustic waves shock and dissipate in the chromosphere, much recent attention has focused on waves associated with the magnetic field. Most observations of propagating waves in magnetic structures are either restricted to the chromosphere and higher layers, or to large features such as sunspots. Here, we present observations of short-period waves propagating along thin, magnetic flux tubes in the lower solar atmosphere, whose transported energy is a promising source of the intensity enhancements of magnetic bright points (MBPs) seen in strong chromospheric lines such as Ca~{\sc ii}~H~\citep{Hasan2008}.

Different modes of oscillation are present in the magnetized solar atmosphere, depending on their compressive/non-compressive and magnetic nature \citep{Edwin1983,Hasan1987,Bogdan2003,Roberts2006,Kato2011}. Such oscillations occur in propagating or standing states \citep{Rosenthal2002,Dorotovic2014}. The magnetic fields act as a guide for magnetohydrodynamics (MHD) waves effectively increasing the connection between different layers of the solar atmosphere. It has been shown that longitudinal acoustic waves in the photosphere propagate along the magnetic field lines and their leakage to the upper atmospheric layers depends on the inclination of the magnetic field \citep{Michalitsanos1973,Bel1977,Suematsu1990,DePontieu2004a,Jefferies2006,Stangalini2011}.

Magnetic waves, excited in a thin flux tube, are generally classified, for simplicity, according to their propagation speeds: (1) Alfv\'{e}n waves propagate with the local Alfv\'{e}n speed, e.g., inside a flux tube, (2) axisymmetric, longitudinal magneto-acoustic waves propagate with a tube speed $c_{T}$ that is smaller than both sound $c_{S}$ and Alfv\'{e}n $c_{A}$ speeds ($c_{T}=c_{S}c_{A}/\sqrt{{c_{S}}^2+{c_{A}}^2}$), and (3) non-axisymmetric, transverse kink waves propagating with a speed between that of the surroundings and $c_{A}$~\citep[e.g.,][]{Spruit1982,Solanki1993,Roberts1997,Cranmer2005,DePontieu2007,Jess2009,Morton2011,Pietarila2011,Mathioudakis2013}. 

Within thicker flux tubes, additional wave modes also propagate: the fast and the slow mode, traveling at speeds faster or slower than the speed of sound, respectively. Thus the speed of fast and slow waves depends on the ratio of $c_{A}/c_{S}$ and on the direction of propagation relative to the field lines. In general, both modes are compressible. The slow mode is closely related to the tube mode (longitudinal magneto-acoustic mode; \citealt{Kato2011}). 
While slow-mode waves dissipate in the chromosphere by forming shocks, fast waves can penetrate into the upper solar atmosphere (due to their higher phase speeds, which reduce the occurrence of shocks). Hence, the latter case, i.e., the fast waves, are of interest for understanding the heating mechanisms of the outer solar atmosphere, while the slow waves may contribute to the heating of the low-to-mid chromosphere.

While each of these waves may propagate along a flux tube, coupling between different modes may also occur~\citep{Roberts2004}. Numerical simulations of wave propagation in the lower solar atmosphere have shown coupling of fast and slow magneto-acoustic waves in regions where the acoustic and Alfv\'{e}n speeds nearly coincide, i.e., where the plasma-$\beta\approx1$ ($\beta \equiv 8 \pi p /B^{2}$;~\citealt{Bogdan2003,Nutto2012}). This level has been called the ``equipartition level''~\citep{Cally2007}, where equipartition between magnetic and thermal energy density is achieved. Away from the equipartition level, the waves with magnetic field-dominated and acoustic natures are decoupled. The modes of such waves in low and high $\beta$ media are summarized in Table~$4$ of~\citet{Bogdan2003}. At the equipartition level, part of the energy contained in the acoustic branch can be channeled to the magnetic branch and vice versa, while, the waves are also partly transmitted through the conversion layer without changing their physical natures, but with exchanging the fast and slow labels. The former and latter cases are normally referred to as ``mode conversion'' and ``mode transmission'', respectively~\citep{Cally2007}. The fraction of energy transmitted or converted depends on the attack angle of the wave (i.e., the angle between the wave and the magnetic field vectors at the transmission/conversion layer) itself, the wavelength, and the width of the conversion layer \citep{Schunker2006,Hansen2009,Stangalini2011}. In addition to the fast and slow waves whose physical nature depends on the level of plasma-$\beta$, (transverse) Alfv\'{e}n waves may propagate in both $\beta>1$ and $\beta<1$ regions~\citep{Rosenthal2002,Bogdan2003}.

Both incompressible (kink) and compressible (e.g., sausage) waves are thought to be excited in magnetic elements through the interaction of flux tubes with surrounding granules~\citep[e.g.,][]{Ulmschneider1991,Hasan1999,Fedun2011,Vigeesh2012}. By exploiting the high spatial resolution images provided by {\sc Sunrise}/IMaX, \citet{Stangalini2013} have reported the interaction between longitudinal and horizontal-velocity oscillations in small magnetic elements in the solar photosphere. Also, propagation of kink waves in small magnetic elements has been shown to be a nonlinear process~\citep{Stangalini2015}.

\citet{Spruit1981a} and \citet{Choudhuri1993b} proposed that rapid, pulse-like motions of the flux-tube footpoints produce kink waves along the tubes that can propagate into the upper solar atmosphere. The energy that such jerky motions can potentially carry may contribute to the heating of the quiet corona~\citep[e.g.,][]{Hasan2000,Hasan2008,Jafarzadeh2013a}. \citet{Cranmer2005} found that the energy of incompressible, transverse kink waves, propagating along flux tubes, are transformed into the volume filling  Alfv\'{e}n waves above the height where individual flux tubes merge (i.e., in the chromosphere). However, according to their model, $95\%$ of such Alfv\'{e}nic waves are reflected at the transition region. Fast magneto-acoustic waves may also couple to Alfv\'{e}n waves at the apex of their refractive path \citep{Felipe2012,Khomenko2012}. This implies a non-homogeneous field or horizontally non-homogeneous density. The propagation of magneto-acoustic and kink waves in small magnetic elements (in the lower solar atmosphere) have been studied in detail in 2D~\citep[e.g.,][]{Khomenko2008} and 3D~\citep[e.g.,][]{Vigeesh2012} simulations.

We note that waves reviewed here propagate in the lower solar atmosphere. Waves in the upper layers of the atmosphere have also been observed, e.g., in coronal loops (\citealt{Aschwanden1999,DeMoortel2002,Wang2003}; see \citealt{Nakariakov2005} for a review), in quiet-Sun with frequencies up to $100$~mHz using the \textit{TRACE} spacecraft~\citep{DeForest2004}, or in spicules from Hinode/SOT with an average frequency of $19$~mHz (\citealt{DePontieu2007,Okamoto2011}, and many others).

In addition, different excitation mechanisms have been proposed or speculated upon for the same wave mode observed at different heights from the solar surface. For instance, the excitation of kink waves observed in the lower solar atmosphere has been attributed to buffeting of the flux tubes by the surrounding granular flows~\citep{Hasan2008}, while, small-scale magnetic reconnection in the chromosphere has been proposed to drive the kink modes detected in the higher atmosphere~\citep{He2009}. \citet{Kato2011} have proposed a mechanism called ``magnetic pumping'', where convective downdrafts around a flux tube pump downflows inside the tube. This mechanism was shown to result in the upward propagation of (slow) magneto-acoustic waves in magnetic flux concentrations~\citep{Kato2016}.

We also note that we have only focused on waves propagation in magnetic elements. In non-magnetized areas, a rich spectrum of waves such as acoustic, gravity, and surface gravity (generating resonant modes of oscillation as $p$-modes, $g$-modes, and $f$-modes below the solar surface, respectively) exists~\citep{Deubner1984,Straus2008}. Propagation of, e.g., acoustic waves in the non-magnetized atmosphere transports considerable energy flux that contributes  to the heating of the chromosphere~\citep{Carlsson1997,Fossum2005,Erdelyi2007,Wedemeyer-Bohm2007,BelloGonzalez2010}.

In this paper, we investigate the propagation of high-frequency waves in the lower solar atmosphere by analyzing both horizontal-displacement oscillations of MBPs and their intensity perturbations at two sampled heights observed at high spatial and temporal resolution with the {\sc Sunrise} balloon-borne solar observatory. We detect high-frequency fast waves, both compressible and incompressible, propagating at the selected MBPs in the lower solar atmosphere.
In Section~\ref{sec-obs} we introduce the data used in this analysis. In Section~\ref{sec-analysis} we describe our analysis methods that retrieve the observational properties of the waves. We provide the results and corresponding interpretations in Section~\ref{sec-statistics}. The results are discussed in Section~\ref{subsec-waves-comparison} and conclusions are drawn in Section~\ref{sec-waves-conclusions}.

\vspace{3mm}
\section{Observations}
\label{sec-obs}

\subsection{Observational Data}
\label{subsec-obsdata}

High spatial and temporal resolution observations of the quiet-Sun disk center at 300~nm (FWHM$\approx5$~nm) and Ca~{\sc ii}~H $396.8$~nm (FWHM$\approx0.18$~nm) were carried out using the {\sc Sunrise} Filter Imager (SuFI;~\citealt{Gandorfer2011}) on board the {\sc Sunrise} balloon-borne solar observatory~\citep{Solanki2010,Barthol2011,Berkefeld2011} on 2009 June 9 (between 01:32~UT and 02:00~UT). Simultaneous full Stokes observations in the magnetically sensitive Fe~{\sc i}~$525.02$~nm line recorded by the Imaging Magnetograph eXperiment (IMaX;~\citealt{Martinez-Pillet2011}) on board {\sc Sunrise} provided the photospheric magnetograms that were used to determine the magnetic properties of the features observed in the other passbands.

\begin{figure}[!tbp]
	\centering
	\includegraphics[width=8.5cm]{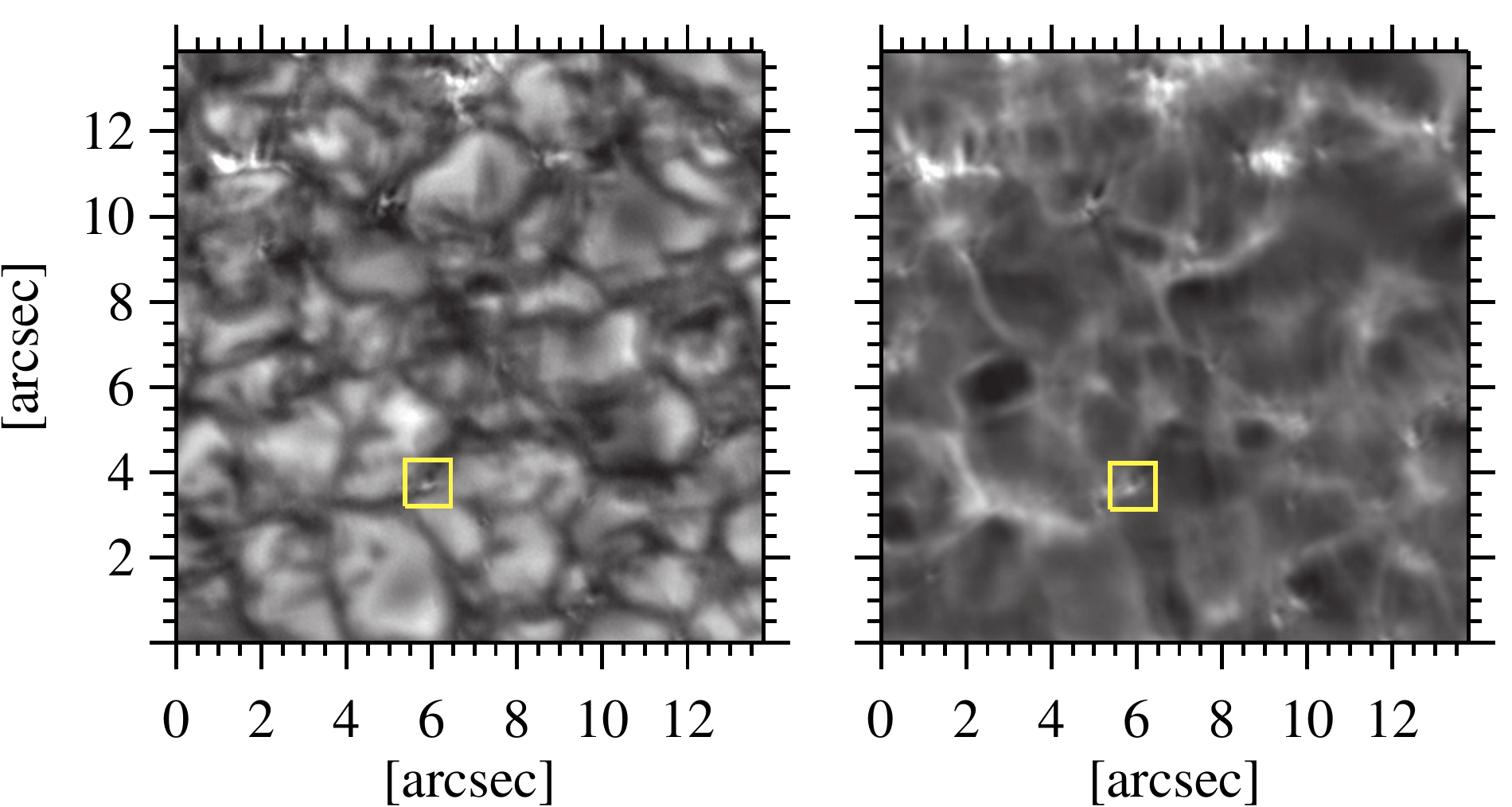}
	\caption{Examples of the {\sc Sunrise}/SuFI images recorded at the 300~nm (left panel) and the Ca~{\sc ii}~H (right panel) passbands. The yellow boxes include a sample magnetic bright point studied in the present work.}
	\label{fig:obs}
\vspace{3mm}
\end{figure}

The seeing-free 300~nm and Ca~{\sc ii}~H image sequences employed in this work share a common field of view of $14\arcsec\times40\arcsec$ and were both recorded with a cadence of $12$~s and a spatial sampling of $\approx0.02$~arcsec/pixel. The recordings at the two wavelengths are offset by $1$~s in time (300~nm images follow Ca~{\sc ii}~H filtergrams). All SuFI images were phase diversity reconstructed with averaged wave front errors (i.e., level 3 data; \citealt{Hirzberger2010,Hirzberger2011}).
Moreover, the IMaX and the two SuFI channels were aligned with sub-pixel accuracy (i.e., better than 14~km), using an approach that utilizes cross-correlation and mean squared deviation techniques applied to common sets of manually selected features (see \citealt{Jafarzadeh2013a} for a description of the alignment procedure).

Figure~\ref{fig:obs} shows a small part of a snapshot in each wavelength. An MBP is marked whose horizontal-displacement fluctuations as well as intensity oscillations are studied here.

\vspace{2mm}
\subsection{Formation Heights}
\label{subsec-formationheights}

We estimate the formation height by two completely independent means. We begin by estimating the heights of formation of the 300~nm band and Ca~{\sc ii}~H line profile by computing their contribution functions (CFs) using the RH radiative transfer code~\citep{Uitenbroek2001}. The code provides the contribution to the emission as a function of height at a certain wavelength by solving both radiative transfer and statistical equilibrium equations in a given atmospheric model. 
The CFs for the 300~nm passband are computed in LTE (Local Thermodynamic Equilibrium) conditions. The calculations are in non-LTE for the Ca~{\sc ii}~H line, taking partial redistribution into account \citep{Magain1986,GrossmannDoerth1988,Uitenbroek1989}. A five-levels Ca~{\sc ii}~H model atom (i.e., with Ca~{\sc ii}~H/K and Ca~{\sc ii} infrared triplet lines; \citealt{Uitenbroek2001}) was used in the latter computations.
Following the discussions of \citet{Jafarzadeh2013a} (hereafter referred to as J13), we use the FALP model atmosphere~\citep{Fontenla1993} to describe the MBPs after convolving the CFs at different wavelengths by the spectral profile of the relevant SuFI filter. For more details of similar implementations, we refer the reader to section $2.1$ in J13. For comparison, the CFs resulting from the FALC model atmosphere (that represents an averaged quite-Sun region) are also determined.

\begin{figure}[!tp]
	\centering
	\includegraphics[width=8.4cm]{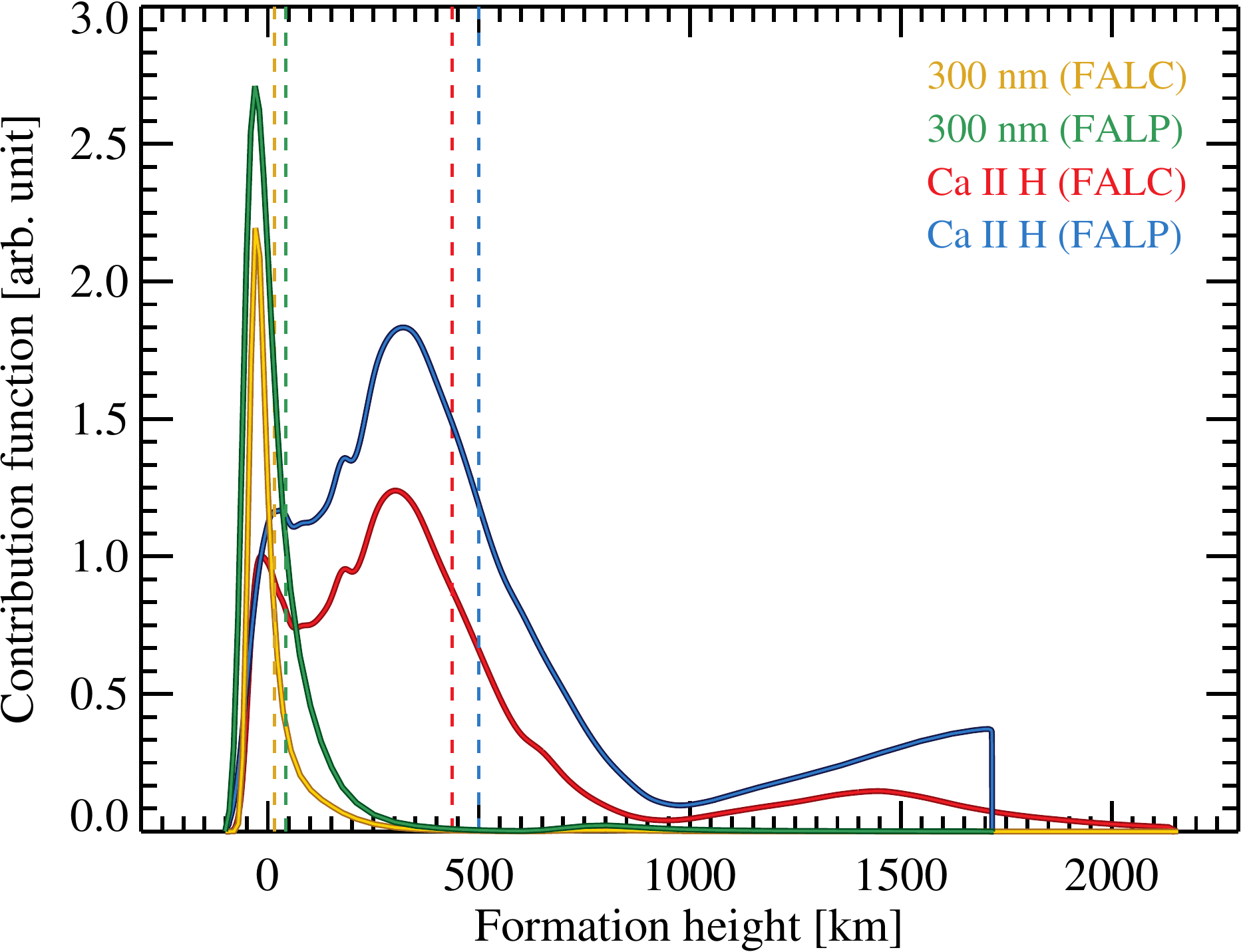}
	\caption{Contribution functions for the {\sc Sunrise}/SuFI 300~nm and Ca~{\sc ii}~H passbands from the RH radiative transfer code, for two atmospheric models (see the text). The vertical dashed lines indicate the corresponding average formation heights.}
	\label{fig:CFs}
\vspace{3mm}
\end{figure}

Plotted in Figure~\ref{fig:CFs} are the CFs for the two passbands and both atmospheric models. We show this figure here, although some of the results can already be found in J13, in order to highlight the difference between the CFs and the average formation heights of the 300~nm and Ca~{\sc ii}~H passbands of {\sc Sunrise}/SuFI. We note that the SuFI Ca~{\sc ii}~H images clearly have a large contribution from the photosphere, with some contribution from the low-to-mid-chromosphere. The vertical dashed lines indicate the corresponding average formation heights. From the mean heights of formation, an average height difference of $\approx450$ is determined between images observed in the 300~nm and the Ca~{\sc ii}~H passbands. The sudden drop of the Ca~{\sc ii}~H CF for the FALP at $\approx1700$~km is due to the rapid temperature increase in the upper chromosphere, leading to the ionization of Ca$^{+}$.

The main uncertainty in these results lies in the choice of the atmospheres. Clearly, the choice of a 1D atmosphere is a gross simplification, because the real Sun shows a rich variety of fine structure and the CFs are affected nonlinearly by changes in the atmospheres. An additional uncertainty is posed by the difference between the CFs and the response functions to individual atmospheric parameters, though response functions in non-LTE are not yet available, so that, this point is somewhat moot.

However, the results shown in Figure~\ref{fig:CFs} can be understood with the help of simple physical considerations. 
A Wilson depression is present inside the flux tubes, so that radiation is emitted from deeper layers inside them. However, we are dealing with the height difference, rather than the absolute heights of formation. The Wilson depression occurs at both heights, making it irrelevant for the difference in formation heights. More important is the run of pressure with height. For a higher temperature inside the magnetic feature, as is required for it to be an MBP, the pressure drops more slowly with height within the magnetic feature than outside it, which increases the difference in formation height, in agreement with the CFs. The formation heights of photospheric spectral lines weakened by a higher temperature, decreases within magnetic elements \citep{Holzreuter2015}. However, this is not the case for the Ca~{\sc ii}~H line core because the increased chromospheric temperature rise within magnetic elements causes the emission peaks around the core of this line to strengthen \citep[e.g.,][]{Skumanich1975,Ayres1986,Solanki1991}, so that the contribution from the chromosphere to this line increases (as indicated by the CF computations shown in Figure~\ref{fig:CFs}). The formation height of the 300~nm continuum also increases slightly when going from quiet-Sun to MBP, as H$^{-}$ opacity increases with temperature. This increase is not large because MBPs are not much brighter than the average quiet-Sun at low heights. Thus we expect the difference in height of formation between the two wavelengths to increase somewhat from quiet-Sun to MBP, which supports the CF computations.

Nonetheless, we check the results obtained from the CFs by employing an entirely independent technique. To this end, we apply an analysis based on the Fourier transform of the intensity oscillations of both, the 300~nm and Ca~{\sc ii}~H time series in a relatively large quiet-Sun FOV (i.e., $14\arcsec\times5\arcsec$). This area was chosen because upon visual inspection it harboured only a relatively small number of MBPs (and no other larger magnetic structures) for the entire time series.

\begin{figure}[!tbp]
	\centering
	\includegraphics[width=8.5cm, trim = 0 0 0 0, clip]{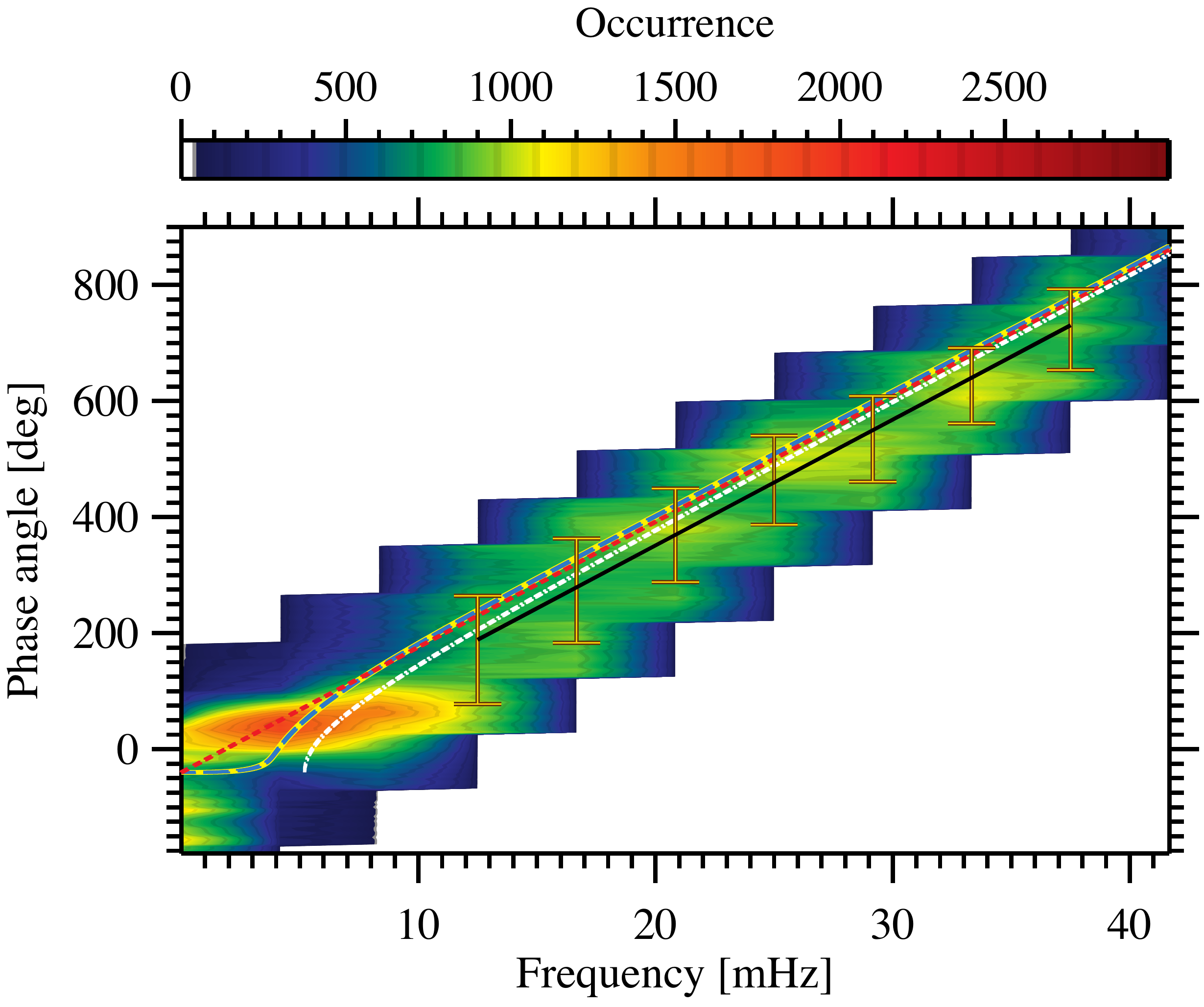}
	\caption{Unwrapped phase spectrum (2D histogram of phase angle as a function of frequency) between the {\sc Sunrise}/SuFI 300~nm and Ca~{\sc ii}~H intensity images in a $14\arcsec\times5\arcsec$ quiet-Sun area; from Fourier transform analysis of individual pixels. The slope, indicated by (black) solid line along with the scatter 1$\sigma$ errorbars, represents wave travel-time of acoustic waves. The fits represent theoretical computations of wave propagation in an isothermal non-stratified atmosphere (red, dashed line), in a stratified atmosphere including gravity (white, dot-dashed line), and in an atmosphere including radiative losses (dark blue, long dashed line).}
	\label{fig:acoustic}
\end{figure}

We perform the Fourier transform on the entire length of the time series, separately for each individual pixel, and compute phase-lags between the time series from their cross spectra. In the averaged quiet-Sun, acoustic waves are expected to be dominant, with $c_{s}\approx7-8$~km\,s$^{-1}$ in the photosphere. Waves propagating in a magnetic element of the chosen quiet-Sun FOV may behave differently. However, their contribution to the obtained phase-lag from the entire FOV is very small, since they only occur in a very small fraction of it. The phase-lag ($\varphi$) can then be converted into a time lag ($\tau$) for each Fourier frequency ($f$) using

\begin{equation}
	\tau = \frac{\varphi }{2\pi  f}\,.
	\label{equ:phase_time}
\end{equation}

Figure~\ref{fig:acoustic} illustrates a 2D histogram of phase versus frequency. It shows the unwrapped phase-diagram, i.e., the phase-diagram that is corrected for phase jumps of $360^{\circ}$. Acoustic waves with frequencies larger than the (corresponding) cut-off frequency can propagate within the atmosphere and indeed we found acoustic waves in the high-frequency range of $4-38$~mHz that are propagating upward. The lower boundary of the frequency range ended at the cut-off frequency of the acoustic waves, the upper boundary at the Nyquist frequency ($\approx42$~mHz). The diagonal ridge outlines a nearly linear trend. We presume that this ridge is due to acoustic waves. The slope of this ridge is given by the best-fit black, solid line. The yellow error bars indicate the standard deviation of the data points along the phase-lag axis. The slope reflects the travel time of the acoustic wave between the two layers and is equal to $57\pm2$~s, which corresponds to a height difference of $430\pm30$~km between the layers of the quiet solar atmosphere sampled by the 300~nm and Ca~{\sc ii}~H passbands. The error comes from the uncertainty in determining the slope and in the sound speed of $7.5\pm0.5$~km\,s$^{-1}$.

For comparison, we have over-plotted the expected phase difference ($\Delta \varphi$) versus frequency ($f$) for linear (vertical) wave propagation between the two layers, based on three different models explained by~\citet{Centeno2006}: 
\begin{enumerate} 
\item Adiabatic propagation in an isothermal non-stratified atmosphere (red, dashed line; Equation~(\ref{equ:phase_time})).
\item In a gravitationally stratified atmosphere (white, dotted--dashed line):

\begin{equation}
	\Delta \varphi = (\Delta h\, \sqrt{\omega ^{2}- {\omega _{ac}}^{2}})/c_{S}\,,
	\label{equ:phase_model2}
\end{equation}
\noindent
where $\omega=2 \pi f$, $\omega_{ac}=\gamma g /2c_{S}$ is the acoustic cut-off frequency, $g=274$~m\,s$^{-2}$ is the gravity (assumed to be constant), and $\gamma=5/3$ is the ratio of the specific heats for adiabatic propagation.

\item Non-adiabatic propagation in a gravitationally stratified atmosphere when radiative losses have been taken into account with Newton's law of cooling (dark blue, long dashed line; originally developed by \citealt{Souffrin1972}):

\begin{equation}
	\Delta \varphi = \Delta h \, \sqrt{\frac{h_{R}+\sqrt{{h_{R}}^{2}+{h_{I}}^{2}}}{2}}\,,
	\label{equ:phase_model3}
\end{equation}
\noindent
where
\begin{equation}
	h_{R}=\frac{\omega ^{2}(1+\omega ^{2}\, {\tau _{R}}^{2}\, \gamma )}{gH_{0}(1+\omega ^{2}\, {\tau _{R}}^{2}\, \gamma ^{2})}-\frac{1}{4{H_{0}}^2}\,,\,
	h_{I}=\frac{\tau _{R}\, \omega ^{3}(\gamma -1)}{gH_{0}(1+\omega ^{2}\, {\tau _{R}}^{2}\, \gamma ^{2})}\,,
	\label{equ:phase_model3_adds}
\end{equation}
\noindent
$H_{0}=({c_{S}}^2)/(\gamma*g)$ is the pressure scale height and $\tau _{R}$ is the radiative time-scale.

\end{enumerate}

We obtained the best fits of the three model curves to the observed trend using $c_{S}=7.5$~km\,s$^{-1}$, $\tau _{R}=12$~s, and $\Delta h \approx 450$~km, which is in agreement with the height difference of $\approx430$~km from the best fit to the data points.

The height difference of $430\pm30$~km between the two atmospheric layers sampled by the {\sc Sunrise}/SuFI Ca~{\sc ii}~H and 300~nm passbands, obtained remarkably from analyzing the acoustic waves in the quiet atmosphere, agrees with that obtained from the CFs (i.e., $\approx420$~km in the quiet-Sun). This gratifying agreement increases our trust in reliability of the formation height differences deduced from the CFs. For the atmosphere inside the magnetic elements, we therefore adopt the value returned by the CFs, 450~km, for this difference, but impose a conservative uncertainty of $\pm100$~km.

\vspace{2mm}
\section{Analysis and a Case Study}
\label{sec-analysis}

We study wave-like phenomena in the lower solar atmosphere by analyzing oscillations in both the horizontal displacement and the intensity in small-scale magnetic features, the intensity oscillations being a tool to investigate the presence of compressive modes in the magnetic elements. Knowledge of the precise position of such elements versus time is a must in such analyses, which require stable (preferably seeing-free) observations. We focus on small magnetic bright points (i.e., MBPs), similar to those studied in J13. 
In short, the MBPs studied here were required to meet the following criteria: they had to (1) be located in an internetwork area, (2) be brighter than the mean intensity of the entire frame, (3) have a magnetic nature (coinciding Stokes $V$ patches with S/N$\geq3$), (4) live longer than five minutes (to avoid short-lived brightenings due to, e.g., high frequency oscillations or reversed granulation at the heights sampled by the Ca~{\sc ii}~H passband; \citealt{Rutten2004,de-Wijn2005}), and (5) be point-like features (with a diameter smaller than 0.3~arcsec). In this way, we ensure that the selected Ca~{\sc ii}~H MBPs differ from non-magnetic H$_{2V}$ or K$_{2V}$ grains, which are mainly due to acoustic waves \citep{Rutten1991,Beck2008}. The availability of polarimetric data enables us to identify magnetic fields and to distinguish between MBPs and other localized brightenings, different from other studies, such as \citet{Keys2013}. Also, the the so-called persistent flashers (whose brightness drops below the detection limit; similar to those shown in J13) are not included in this study.

Consequently, the accurate locations as well as the intensity of such MBPs at any given time are determined using the same algorithm as described in detail in J13. We search for MBPs whose horizontal displacement and intensity show an oscillatory behavior, such as the examples presented in J13. In order to facilitate the precise localization and tracking of the MBPs, noise and extended solar brightenings were eliminated from both sets of images (i.e., granules from the 300~nm and other brigthenings due to, e.g., shock waves from the Ca~{\sc ii}~H filtergrams). This approach returns the locations of the MBPs with an accuracy better than 0.5~pixel ($\approx14~km$). See section 3.1 of J13 for details on the image processing, detection and tracking algorithms employed here.

In order to determine whether any of the detected waves propagate within the solar atmosphere (and to measure the speed of propagation), we need to simultaneously trace horizontal displacements and intensity oscillations of the same magnetic elements at two atmospheric heights.

Thus, we found seven MBPs whose trajectories could be precisely tracked in both the 300~nm and the Ca~{\sc ii}~H image sequences for a sufficiently long time, i.e., longer than five minutes. This latter criterion improves the frequency resolution in the observed oscillations The amplitudes of the oscillations in the seven MBPs at the two atmospheric heights, along with the periods and phase angles of the detected waves, are summarized in Table~\ref{table:stat}.

The number of MBPs investigated here is lower than the {\sc Sunrise} Ca~{\sc ii}~H BPs (with lifetimes longer than five minutes) studied in J13 because of the relatively low contrast of the 300~nm filtergrams compared to the Ca~{\sc ii}~H images~\citep{Riethmuller2010}. We dropped all of those MBPs that could not be precisely tracked in both filtergrams over their entire lifetimes. By restricting ourselves to the smallest MBPs, we may have discarded the longer lived ones (if larger MBPs have longer lifetimes, which, however, is not the case according to \citealt{Anusha2017} if one includes splitting and merging among the causes of death). Due to the careful choice of MBPs, coupled with the seeing-free high-resolution data, we expect to get reliable results on the waves within the MBPs. Nonetheless, given the small number of MBPs in our sample, the results we obtained cannot be taken to represent the properties of quiet-Sun MBPs in general.

We note that, although uncertainties introduced by the locating algorithm as well as instrumental vibration-induced image jitter may bias the final analysis, their effects are much smaller than the motions of the MBPs under study. \citet{Jafarzadeh2013a} found an average uncertainty of $0.02$~km\,s$^{-1}$ in the determined horizontal velocities of similar MBPs. This is much lower than the mean amplitudes of the horizontal velocities of both 300~nm and Ca~{\sc ii}~H MBPs (i.e., 1.8~km\,s$^{-1}$ and 1.2~km\,s$^{-1}$, respectively; see Table~\ref{table:stat}) in the present study. The power spectra of the residual image jitter (in both horizontal and vertical directions) measured for the {\sc Sunrise} observatory show a frequency range of about $10-150~Hz$ \citep{Berkefeld2011}, which is much higher than the sub-Hertz frequencies of the waves under study (see Section~\ref{sec-statistics}). Hence, the effect of instrumental vibration-induced jitter is negligible in our analysis, except for smearing the images. Moreover, we observe clear phase-lags between the oscillations observed in the two heights (see Section~\ref{subsec-results}) that cannot be due to jitter or random motions.

The maximum intensity over all pixels of an MBP at any given time is measured as its intensity at that time step. \citet{Jafarzadeh2013a} obtained an average photon noise of $0.01~\langle I_{Ca} \rangle$ in the Ca~{\sc ii}~H images as a mean uncertainty in determining the intensity. This noise level is much smaller (by a factor of $15$) than the mean amplitude of the Ca~{\sc ii}~H intensity oscillations.

The $1$~s time difference between the two time series (i.e., 300~nm and Ca~{\sc ii}~H image sequences) is corrected for by adding the corresponding phase angle to the final results.

\vspace{3mm}
\subsection{Wavelet Transform}
\label{subsec-wavelet}

We perform a wavelet analysis~\citep{Daubechies1990,Torrence1998} in order to simultaneously localize the spectral power in both time and frequency domains. The wavelet transform is especially suitable for searching transient oscillations and for studying wave propagation within the solar atmosphere in the presence of short-lived features and waves~\citep{Baudin1994,Baudin1996,Bloomfield2004,McIntosh2004,Jess2007}.

The wavelet ($W$) is defined as the convolution of a time series with a ``mother'' function that is a window (envelope) whose variable width enables the analysis to capture both low/high frequencies and long/short durations simultaneously. We employ the Morlet mother function with a dimensionless frequency $\omega_{0}=6$ using a wavelet algorithm developed by~\citet{Grinsted2004}. This function satisfies the balance between frequency and time localization~\citep{Farge1992} and hence is suitable for investigating the propagation of waves with different ranges of frequencies. The Morlet function is a complex wavelet, consisting of a plane wave modulated by a Gaussian window, described as

\begin{equation}
\psi _{0}\left ( \eta  \right ) = \pi^{-1/4}\, e^{i\omega_{0}\eta }\, e^{-\eta^{2}/2}\,,
	\label{equ:morlet}
\end{equation}
\noindent
where $\eta=s\,t$ is a dimensionless time that stretches the wavelet in time $t$ by changing its scale $s$.

The wavelet power spectrum of a time series is defined as $|W(s)|^{2}$. The cross-spectrum (or cross wavelet power spectrum) of two time series is then determined by multiplying the wavelet power spectrum of a time series with the complex conjugate of the other one. Furthermore, interaction between the two time series can be examined using a bivariate framework called wavelet coherence, which is the square of the cross-spectrum normalized by the individual power spectra~\citep{Grinsted2004}. The coherence level varies between zero and one representing incoherent and coherent oscillations, respectively.

\begin{table*}[!tp]
\begin{center}
\caption{Properties of Oscillations and of Detected Waves in Magnetic Bright Points (MBPs) Studied in the Present Work}
\label{table:stat}
\setlength{\tabcolsep}{0.45em}   
\renewcommand{\arraystretch}{1.6}                 
\begin{tabular}{l c c c c c c c c c c c c c c c c c c c}
\Xhline{0.8pt}
\\[-2em]
\Xhline{0.8pt}
\\[-1.4em]
\multicolumn{2}{c}{MBP} & Lifetime\tablenotemark{a} & \multicolumn{8}{c}{Displacement Oscillations} && \multicolumn{8}{c}{Intensity Oscillation} \\
\cline{4-11} \cline{13-20}
\multicolumn{2}{c}{ } & (s) & \multicolumn{2}{c}{amplitude (km\,s$^{-1}$)} && \multicolumn{2}{c}{period\tablenotemark{b} (s)} && \multicolumn{2}{c}{phase angle\tablenotemark{b} (deg)} && \multicolumn{2}{c}{amplitude (arb.\tablenotemark{c})} && \multicolumn{2}{c}{period\tablenotemark{b} (s)} && \multicolumn{2}{c}{phase angle\tablenotemark{b} (deg)} \\
\cline{1-2} \cline{4-5} \cline{7-8} \cline{10-11} \cline{13-14} \cline{16-17} \cline{19-20}
NO. & Passband & & mean & peak && mean & median && mean & median && mean & peak && mean & median && mean & median \\
\Xhline{0.8pt}
\multicolumn{1}{l|}{\multirow{2}{*}{1}} & 300~nm & {\multirow{2}{*}{528}} & 
$1.3\pm0.2$\tablenotemark{d} & 3.7 && {\multirow{2}{*}{$47\pm1$}} & {\multirow{2}{*}{45}} && {\multirow{2}{*}{$130\pm5$}} & {\multirow{2}{*}{114}} &&
$25\pm3$ & 72 && {\multirow{2}{*}{$128\pm13$}} & {\multirow{2}{*}{118}} && {\multirow{2}{*}{$44\pm2$}} & {\multirow{2}{*}{40}} \\
\multicolumn{1}{l|}{} & Ca~{\sc ii}~H & & 
$0.9\pm0.1$ & 2.1 && & && & && 
$242\pm47$ & 768 && & && & \\
\Xhline{0.8pt}
\multicolumn{1}{l|}{\multirow{2}{*}{2}} & 300~nm & {\multirow{2}{*}{552}} & 
$2.3\pm0.3$ & 7.8 && {\multirow{2}{*}{$57\pm1$}} & {\multirow{2}{*}{59}} && {\multirow{2}{*}{$65\pm1$}} & {\multirow{2}{*}{67}} &&
$53\pm7$ & 99 && {\multirow{2}{*}{$116\pm11$}} & {\multirow{2}{*}{140}} && {\multirow{2}{*}{$71\pm5$}} & {\multirow{2}{*}{79}} \\
\multicolumn{1}{l|}{} & Ca~{\sc ii}~H & & 
$1.0\pm0.1$ & 2.6 && & && & && 
$463\pm73$ & 1306 && & && & \\
\Xhline{0.8pt}
\multicolumn{1}{l|}{\multirow{2}{*}{3}} & 300~nm & {\multirow{2}{*}{408}} & 
$1.9\pm0.3$ & 6.8 && {\multirow{2}{*}{$48\pm1$}} & {\multirow{2}{*}{50}} && {\multirow{2}{*}{$89\pm15$}} & {\multirow{2}{*}{106}} &&
$79\pm12$ & 189 && {\multirow{2}{*}{$125\pm8$}} & {\multirow{2}{*}{125}} && {\multirow{2}{*}{$30\pm1$}} & {\multirow{2}{*}{31}} \\
\multicolumn{1}{l|}{} & Ca~{\sc ii}~H & & 
$1.3\pm0.2$ & 3.4 && & && & && 
$414\pm69$ & 882 && & && & \\
\Xhline{0.8pt}
\multicolumn{1}{l|}{\multirow{2}{*}{4}} & 300~nm & {\multirow{2}{*}{624}} & 
$2.1\pm0.2$ & 6.5 && {\multirow{2}{*}{$49\pm1$}} & {\multirow{2}{*}{51}} && {\multirow{2}{*}{$86\pm2$}} & {\multirow{2}{*}{90}} &&
$64\pm10$ & 191 && {\multirow{2}{*}{$114\pm3$}} & {\multirow{2}{*}{99}} && {\multirow{2}{*}{$29\pm2$}} & {\multirow{2}{*}{37}} \\
\multicolumn{1}{l|}{} & Ca~{\sc ii}~H & & 
$1.6\pm0.2$ & 7.2 && & && & && 
$652\pm64$ & 978 && & && & \\
\Xhline{0.8pt}
\multicolumn{1}{l|}{\multirow{2}{*}{5}} & 300~nm & {\multirow{2}{*}{480}} & 
$2.0\pm0.3$ & 5.9 && {\multirow{2}{*}{$53\pm1$}} & {\multirow{2}{*}{53}} && {\multirow{2}{*}{$112\pm2$}} & {\multirow{2}{*}{109}} &&
$69\pm11$ & 229 && {\multirow{2}{*}{$73\pm9$}} & {\multirow{2}{*}{92}} && {\multirow{2}{*}{$18\pm5$}} & {\multirow{2}{*}{15}} \\
\multicolumn{1}{l|}{} & Ca~{\sc ii}~H & & 
$1.2\pm0.2$ & 3.3 && & && & && 
$274\pm38$ & 705 && & && & \\
\Xhline{0.8pt}
\multicolumn{1}{l|}{\multirow{2}{*}{6}} & 300~nm & {\multirow{2}{*}{396}} & 
$1.2\pm0.2$ & 2.6 && {\multirow{2}{*}{$54\pm1$}} & {\multirow{2}{*}{53}} && {\multirow{2}{*}{$98\pm3$}} & {\multirow{2}{*}{96}} &&
$24\pm3$ & 43 && {\multirow{2}{*}{$92\pm9$}} & {\multirow{2}{*}{109}} && {\multirow{2}{*}{$42\pm9$}} & {\multirow{2}{*}{45}} \\
\multicolumn{1}{l|}{} & Ca~{\sc ii}~H & & 
$1.1\pm0.2$ & 3.2 && & && & && 
$199\pm45$ & 840 && & && & \\
\Xhline{0.8pt}
\multicolumn{1}{l|}{\multirow{2}{*}{7}} & 300~nm & {\multirow{2}{*}{420}} & 
$1.7\pm0.2$ & 3.6 && {\multirow{2}{*}{$47\pm1$}} & {\multirow{2}{*}{47}} && {\multirow{2}{*}{$75\pm9$}} & {\multirow{2}{*}{78}} &&
$62\pm7$ & 101 && {\multirow{2}{*}{$103\pm8$}} & {\multirow{2}{*}{106}} && {\multirow{2}{*}{$33\pm1$}} & {\multirow{2}{*}{35}} \\
\multicolumn{1}{l|}{} & Ca~{\sc ii}~H & & 
$1.6\pm0.2$ & 4.3 && & && & && 
$380\pm70$ & 1270 && & && & \\
\Xhline{0.8pt}
\vspace{-6mm}
\end{tabular}
\end{center}
\hspace{1mm}
\textbf{Notes.}
\vspace{0.7mm}
\footnotetext[1]{ The statistical correction to the lifetimes of the MBPs applied by J13 on the lifetime distribution is not introduced here.}
\footnotetext[2]{ Periods and phase angles are computed from the wavelet coherence, as long as they are located outside the COI and within contours of 95\% confidence.}
\footnotetext[3]{ Arbitrary unit.}
\footnotetext[4]{ All errors in the table represent uncertainties in the mean values.}
\end{table*}

\vspace*{6mm}
The cross-spectrum highlights time-frequency areas with high common power in the two time series, whereas the wavelet coherence detects regions in a time-frequency domain where the examined time series co-move, but do not necessarily possess a strong common power. Hence, while the cross-spectrum can provide sufficient information on oscillatory behaviors in a localized medium (by representing a local co-variance between two time series), the wavelet coherence is also needed for finding co-movements between perturbations at different regions (heights) in the solar atmosphere. 

Since the wavelet transform of a time series with a finite length has edge artifacts (the wavelet is not totally localized in time), a ``cone of influence'' (COI) is introduced. Thus, the edge effects inside the COI cannot be ignored. Following ~\citet{Grinsted2004}, the COI is defined as regions where the wavelet power spectra from a discontinuity at the edge has reached e$^{-2}$ of the value at the edge.

Finally, a phase difference between a pair of time series provides information on delays in the oscillation (i.e., on wave propagation). These phase-lags are estimated from the complex and real arguments of the cross spectra.

\vspace{2mm}
\subsection{A Case Study: Wavelet Analysis of an MBP}
\label{subsec-results}

In the present study, we compute wavelet power spectra of four time series: the horizontal displacement and the intensity oscillations at two sampled heights (i.e., the two wavelength bands) for a given MBP. This is done for all MBPs, but here we discuss the results in greater detail for one example. Then, we determine cross power and wavelet coherence between the power spectra of the two time series of a given type of oscillation sampled corresponding to the two atmospheric layers.

\begin{figure*}[tbp]
	\centering
	\includegraphics[width=0.95\textwidth, trim = 0 0 0 0, clip]{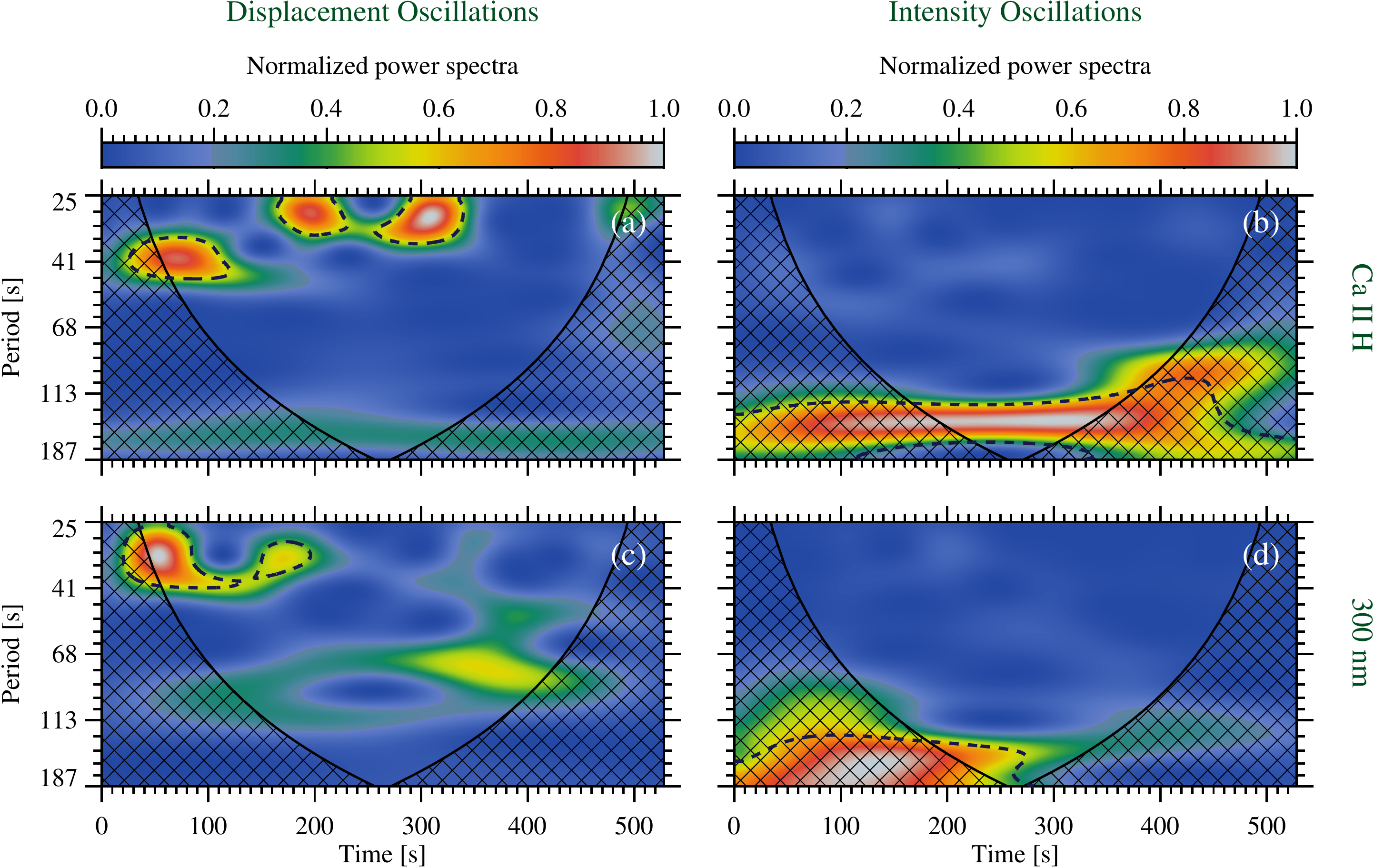}
	\caption{Wavelet power spectra of the horizontal displacement (a), (c) and the intensity (b), (d) of the sample magnetic bright point marked in Figure~\ref{fig:obs}, as observed with time series of images in the Ca~{\sc ii}~H passband (a), (b), and in the $300$~nm passband (c), (d).
	The cross-hatched area indicates the cone-of-influence (COI), representing time-period regions that are subject to edge effects (see main text). The black (dashed) contours mark the 95\% confidence level.
	}
	\label{fig:wps}
\vspace*{6mm}
\end{figure*}

\vspace*{6mm}
Figure~\ref{fig:wps} illustrates the wavelet power spectra of the horizontal-displacement perturbations and the intensity oscillations observed in both the Ca~{\sc ii}~H and 300~nm passbands for the MBP highlighted in Figure~\ref{fig:obs}. The background colors represent the power normalized to its maximum value. The cross-hatched areas show the COI, which defines regions that are subject to the edge effect. Hence, only the computed values outside the COI are considered for the phase analysis. The black contours indicate the 95\% confidence level.

The MBP under study has a lifetime of $528\pm9$~s (for details on determining the lifetime and its uncertainty of similar MBPs we refer the reader to Section~$3.2.2$ of J13). The wavelet analysis of both types of oscillations reveals periodicities with a wide range of values ($25-187$~s) for this magnetic element.
While the intensity perturbation of the MBP has much of the power in the period range of $\approx90-190$~s, its horizontal-displacement oscillates within a higher frequency range (periods of 25-50~s).
Note that the lower limit of the period range is slightly larger than the shortest detectable period of 24~s that corresponds to the Nyquist frequency of our image sequences. 

Figures~\ref{fig:wps}(a) and \ref{fig:wps}(c) show several power peaks, indicating the occurrence of horizontal-displacement oscillations (observed at the heights sampled by Ca~{\sc ii}~H and 300~nm, respectively) with periods changing over time. By visual inspection of these power maps, we can find two high power patches in the period range $25-50$~s and within the time interval 20-250~s. These seem to occur in Figure~\ref{fig:wps}(a) (corresponding to the oscillations observed in Ca~{\sc ii}~H) after equivalent features in Figure~\ref{fig:wps}(c) (representing the oscillations seen in the 300~nm bandpass). Another patch with similar behavior (but with weaker power) in the same period range at around time 500~s is also observed.
The power patches in intensity oscillations, displayed in Figures~\ref{fig:wps}(b) and \ref{fig:wps}(d), show that much of the power is concentrated at longer periods than that of the horizontal-velocity perturbations.
We note that in all plots only small areas outside the COI correspond to high power values. However, they include patches of sufficiently large power for the oscillatory motions to be real (i.e. lying over the 95\% confidence level). Moreover, we are interested in the correlations of the oscillations between the two sampled heights. 

\begin{figure*}[!thp]
	\centering
	\includegraphics[width=0.95\textwidth, trim = 0 0 0 0, clip]{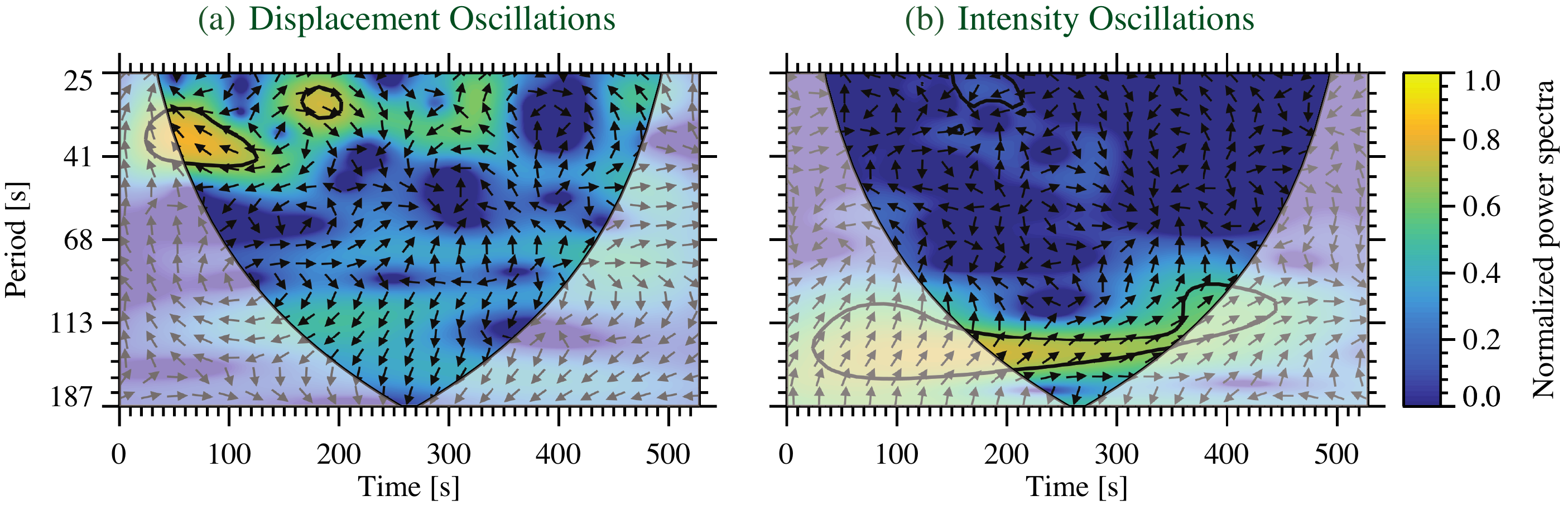}
	\caption{Cross wavelet spectrum between the time series of images in the 300~nm and in the Ca~{\sc ii}~H passband (see Figure~\ref{fig:wps}): (a) for the horizontal displacement, (b) for the intensity.
	 The arrows indicate the phase-lag between the two time series observed at the two sampled heights (with in-phase oscillations marked by arrows pointing right and Ca~{\sc ii}~H following 300~nm by $90^{\circ}$ depicted by arrows pointing straight up). The 95\% confidence level is indicated by black contours. The cone of influence (COI) lies within the bleached or light shaded region.
	 }
	\label{fig:cross}
\vspace*{2mm}
\end{figure*}

The correlations between the power spectra shown in Figures~\ref{fig:wps}(a) and \ref{fig:wps}(c) and those in Figures~\ref{fig:wps}(b) and \ref{fig:wps}(d) are illustrated in Figures~\ref{fig:cross}(a) and \ref{fig:cross}(b), respectively. The cross spectra represent the regions of common high powers in the time-frequency domain. The contours mark the 95\% confidence level. The COI is shown as areas with bleached colors. 

\begin{figure*}[!thp]
	\centering
	\includegraphics[width=0.95\textwidth, trim = 0 0 0 0, clip]{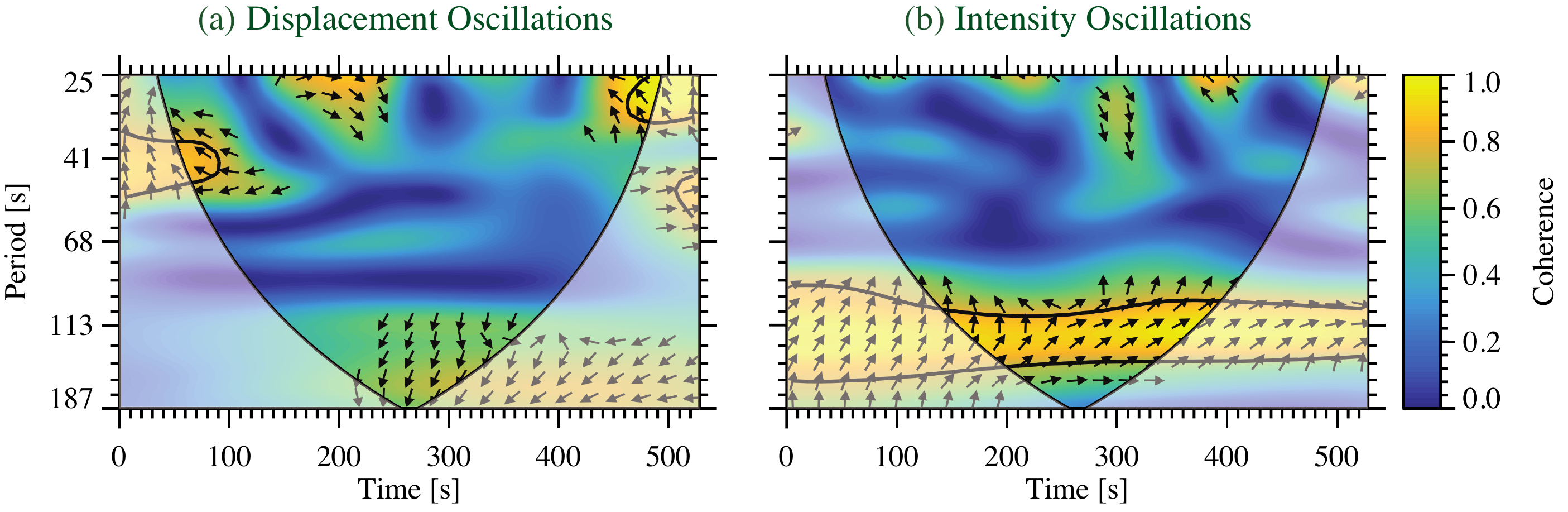}
	\caption{Wavelet coherence between fluctuations in the 300~nm and the Ca~{\sc ii}~H image sequences (see Figures~\ref{fig:wps} and \ref{fig:cross}) of: (a) the horizontal displacement, (b) the intensity.
	 The arrows are the same as in Figure~\ref{fig:cross}, but for clarity, are only depicted in areas with a coherence level exceeding 0.5. The contours indicate a 95\% confidence level as determined using Monte Carlo simulations. The cone of influence (COI) is indicated by the bleached colors.
	 }
	\label{fig:cohr}
\vspace*{6mm}
\end{figure*}

The phase angles between the two atmospheric layers can be deduced from the small arrows in both panels of Figure~\ref{fig:cross}. Arrows pointing to the right and to the left show in-phase and anti-phase oscillations between the two layers, respectively (i.e., representing standing waves). Arrows pointing straight down indicate that the oscillations in Ca~{\sc ii}~H lead the oscillations in 300~nm images by $90^{\circ}$, which would imply downward propagating waves.

The cross-spectrum of displacement oscillations clearly shows the two common patches of high power in Figures~\ref{fig:wps}(a) and \ref{fig:wps}(c) mentioned earlier. However, the arrows indicate that only the leftmost patch corresponds to upwardly propagating waves. The rightmost patch in Figure~\ref{fig:cross}(a) represents waves moving downward in the solar atmosphere. The intensity perturbations of common high power and long periods indicate upwardly propagating waves of a small phase shift.

The wavelet coherence of the two types of oscillations is shown in Figure~\ref{fig:cohr}. To ensure the reliability of the results, only values within contours of 95\% confidence are included for the phase analysis. The confidence level is computed using a Monte Carlo method, because the statistical behavior of the wavelet coherence is unknown~\citep{Torrence1998,Grinsted2004}. We note that relatively large values of wavelet coherence do not necessarily intimate significant co-movement between the two time series. The wavelet coherence attains large, significant values in areas which do not necessarily represent high power patches in the individual wavelet power spectra, in particular for the horizontal displacements. Therefore, the individual power spectra are not a particularly good guide to the presence of correlated oscillations at the two heights; the wavelet coherence provides better information on wave propagation.

The phase angles are also indicated on Figures~\ref{fig:cohr}(a) and \ref{fig:cohr}(b), but for simplicity, only for areas with a coherence value larger than 0.5. 
In both panels,~\ref{fig:cohr}a and \ref{fig:cohr}b, and in all regions within contours of 95\% confidence that are outside the COI, the arrows tend to point upward (i.e., positive phase-angle). This indicates that the oscillations observed in the Ca~{\sc ii}~H passband follow the ones seen in the 300~nm filtergrams, meaning that both types of waves propagate upward in the magnetic element under study.

Different angles of the upward pointing arrows represent different phase angles between the oscillations in the two layers which can be converted to time lags $\tau$ (i.e., the wave travel time) for specific frequencies using Equation~\ref{equ:phase_time}.

We compute the time lags for all points of the wavelet-coherence map (e.g., Figure~\ref{fig:cohr}) which are located outside the COI and have a confidence level of 95\% or higher.

The propagation speed of the waves ($C_{w}$) at a given frequency can then be calculated from its corresponding time lag ($\tau$) and the height difference ($h$) between the two atmospheric layers, i.e., $C_{w}=h/ \tau$.

\vspace{2mm}
\section{Statistics and Wave Properties}
\label{sec-statistics}

Table~\ref{table:stat} summarizes properties of the detected oscillations and waves in the individual seven MBPs studied here. The lifetime represents the duration along which an MBP is simultaneously observed in both the 300~nm and Ca~{\sc ii}~H images.
The amplitudes of both types of oscillations, i.e., fluctuations in displacement (horizontal velocity) and in intensity of the MBPs, are also provided. In agreement with J13, the MBPs under study turned out to have high-velocity excursions (i.e., large horizontal-velocity amplitudes) over the course of their lifetimes. Such rapid ``pulses'' have been shown to excite kink waves in magnetic elements \citep{Spruit1981a,Choudhuri1993a}. The MBPs detected in the 300~nm images have, on average, larger horizontal-velocity amplitudes than their Ca~{\sc ii}~H counterparts.
The mean and median values of period and of phase-angle for each oscillation observed in the individual MBPs are obtained from their computed wavelet coherence, as long as they are located outside the COI and within contours of 95\% confidence (see Section~\ref{subsec-results}).

\begin{figure*}[!tp]
	\centering
	\includegraphics[width=0.95\textwidth, trim = 0 0 0 0, clip]{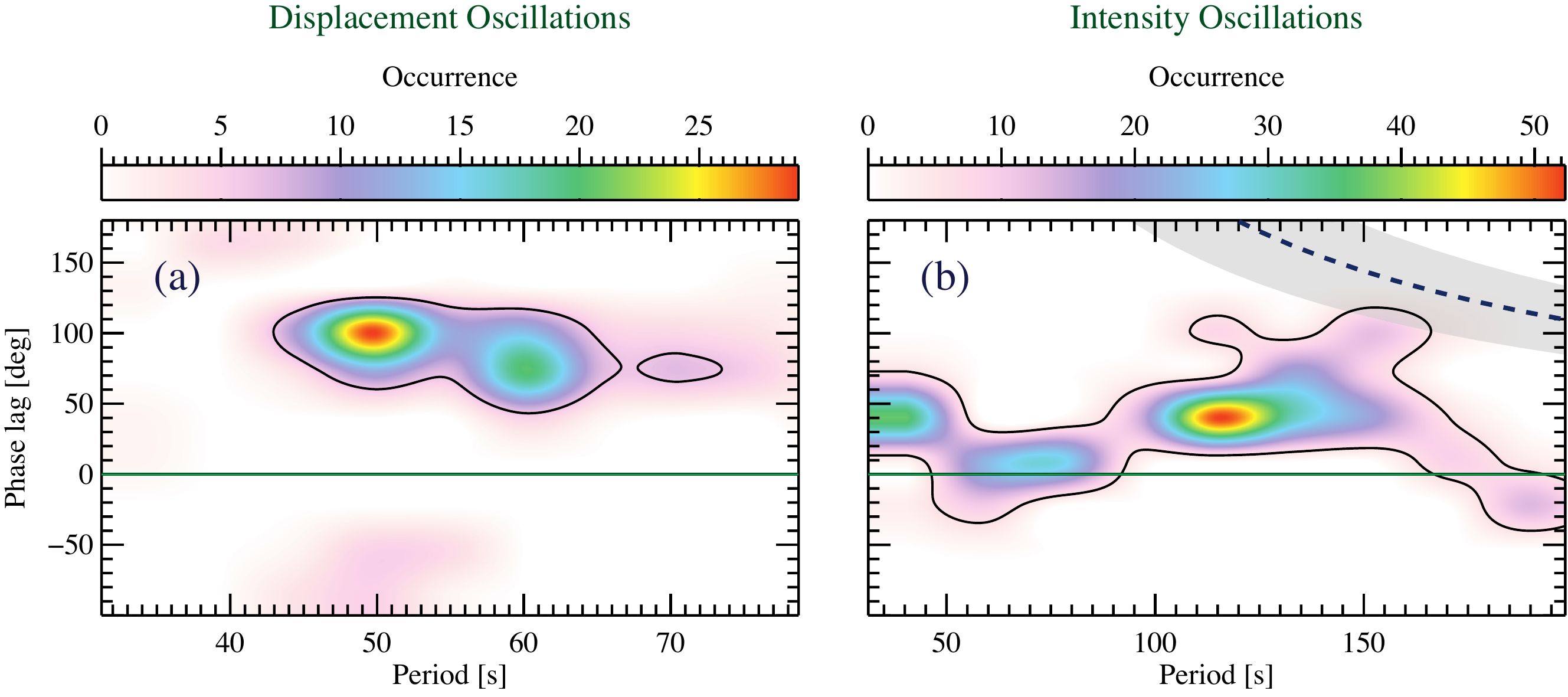}
	\caption{Phase diagram (2D histogram of phase-lag versus period) of the horizontal-displacement oscillations (a) and the intensity perturbations (b) in small magnetic bright points observed in the passbands of 300~nm and Ca~{\sc ii}~H. Positive phase-lags represent upward propagation in the solar atmosphere. The dashed curve, surrounded by the shaded area, represents the dispersion relation of acoustic waves for a height difference of $450\pm100$~km between the formation heights of the two passbands. The solid-line contours separate the statistically reliable regions from extreme outliers (see the text). The green line identifies zero phase difference.
	}
	\label{fig:2dstat}
\vspace*{4mm}
\end{figure*}

Figures~\ref{fig:2dstat}(a) and \ref{fig:2dstat}(b) are 2D histograms of phase-angle versus period from the wavelet coherence of all seven investigated MBPs of the horizontal velocity and the intensity oscillations, respectively. All phase-angle values obtained from all seven analyzed MBPs enter Figures~\ref{fig:2dstat}(a) and \ref{fig:2dstat}(b), as long as they are located outside the COI and within contours of 95\% confidence. It turns out that all these phase angles have a coherence value exceeding $0.7$. Examples are the phase-angles obtained from the wavelet coherence spectra satisfying the above criteria in Figures~\ref{fig:cohr}(a) and \ref{fig:cohr}(b).

The background colors in Figures~\ref{fig:2dstat}(a) and \ref{fig:2dstat}(b) represent the sample density. For robust statistics, unbiased by outliers with a small probability of occurrence (due to, e.g., spurious coherence between the two oscillations), the extreme outliers, i.e., values lying outside the clusters of most of the other data points, are determined. We separate concentrated regions from the extreme outliers using the Grubbs statistical significance test~\citep{Grubbs1969}. This test estimates a criterion corresponding to the largest deviation from the mean value in units of the standard deviation, $\sigma$, for the data points in the 2D histograms. Regions with a number density smaller than this criterion are considered to host the outliers whose small probability of occurrence compared to the rich cluster of other data points may lead to biased interpretations~\citep{Press2007}. The solid-line contours in Figures~\ref{fig:2dstat}(a) and \ref{fig:2dstat}(b) indicate the estimated Grubbs' criteria of $6.3\sigma$ and $5.9\sigma$, respectively, from the mean values of the concentrated regions. 
The contours include 1032 and 2274 individual data points obtained from the wavelet coherence of the seven MBPs, for the horizontal displacement and intensity oscillations, respectively.

We find short-period (high-frequency) oscillations with periods (1) between 43~s and 74~s for horizontal displacements and (2) between $31$~s and $197$~s for intensity oscillations. The phase-angle distributions show almost only positive values, meaning the propagation of wave-like phenomena from the height sampled by the 300~nm passband (i.e., the lower layer) toward the height sampled by the Ca~{\sc ii}~H filter. However, about $16\%$ of all occurrences seen in the intensity oscillations have a negative phase-angle, which can represent a downward propagation of the perturbations. Also, about 1\% of the data points related to the intensity oscillations have a zero phase lag (standing mode). 

The wave travel times corresponding to the highest peaks of the 2D histograms in Figures~\ref{fig:2dstat}(a) and \ref{fig:2dstat}(b) were computed (using Equation~(\ref{equ:phase_time})) as $\approx15$~s and $\approx14$~s for the horizontal displacement and intensity oscillations, respectively.

Both phase diagrams, particularly that of the intensity oscillations, show patches of high occurrences at distinct period ranges. We cannot, however, verify whether these separated regions represent distinct dispersion relations due to, e.g., different waves with different modes or natures.

We perform a test to inspect whether the time delays corresponding to the isolated islands in Figure~\ref{fig:2dstat}(b) are compatible with what we expect from acoustic waves. We follow \citet{Lawrence2011} and \citet{Lawrence2012} and assume that the trend connecting the peaks in Figure~\ref{fig:2dstat}(b), or a part of that, represents the acoustic dispersion relation. Consequently, the height difference between the two layers can be estimated using the wave travel time $\tau$ (using Equation~(\ref{equ:phase_time})) and a sound speed of $c_{S}\approx 7-8$~km\,s$^{-1}$. In this case, we obtain $\Delta h \approx 20-120$~km, a height difference that is small compared to the $\approx 450$~km estimated from radiative transfer (see Section~\ref{subsec-formationheights}). A somewhat larger $c_{S}$, which may be found in a hot magnetic element, is not able to resolve this discrepancy. For comparison, the dispersion relation expected for acoustic waves traveling between the two atmospheric layers (with a height difference of $450$~km) is also plotted in Figure~\ref{fig:2dstat}(b) (the dashed curve). To illustrate the effect of any uncertainty in the formation height of Ca~{\sc ii}~H, the gray shaded area around the dashed curve shows how strongly the latter dispersion relation would change for a range of $\pm100$~km around the 450~km height difference. The acoustic phase-lag is obviously much larger than that observed in the MBPs at all frequencies. This means that either the waves in the magnetic features propagate much faster than sound, or the comparable height difference determined using two independent approaches (i.e., from the CFs, and from the Fourier analysis of acoustic waves; see Section~\ref{subsec-formationheights}) are wrong. Alternatively, the short wave travel times could represent wave fronts traveling obliquely to the line of sight~\citep{Nutto2012}. This would imply observations of phase speeds and not true propagation speeds of waves (see Section~\ref{subsec-waves-comparison}).

Since the difference in formation height between 300~nm and Ca~{\sc ii}~H increases slightly when going to a hotter atmosphere (see Section~\ref{subsec-formationheights}), we conclude that the problem lies in assigning the intensity oscillations in the MBPs to acoustic or slow-mode magneto-acoustic waves.

Using a height difference of $\approx430\pm30$~km and the wave travel times determined from the 2D histograms shown in Figure~\ref{fig:2dstat}, we estimate an upward propagation speed of $29\pm2$~km\,s$^{-1}$ for the horizontal displacements and a value of $31\pm2$~km\,s$^{-1}$ for the intensity oscillations. If one identifies these velocities as kink or Alfv\'{e}n speeds ($c_{K}$ or $c_{A}$, respectively), then, using

\begin{equation}
 	B_{0} = c_{K} \left (\sqrt{4\pi \left ( \rho _{0}+\rho _{e} \right )} \right )\,,
	\label{equ:B_kink}
\end{equation}
and
\begin{equation}
 	B_{0} = c_{A} \left (\sqrt{4\pi \, \rho _{0}} \right )\,,
	\label{equ:B_alfven}
\end{equation}
\noindent
we can obtain a rough estimate of the field strength. In Equations (\ref{equ:B_kink}) and (\ref{equ:B_alfven}), $B_0$ is the field strength in the flux tube, $\rho _{0}\approx2\times10^{-8}$~g/cm$^{3}$ and $\rho _{e}\approx1\times10^{-7}$~g/cm$^{3}$ are the average gas densities (averaged over the atmosphere between the two sampled heights, from FALP and FALC models) inside and outside magnetic elements, respectively. Although the Alfv\'{e}n wave is incompressible, for simplicity, we use the Alfv\'{e}n speed as a lower limit for the speed of the fast magnetosonic wave (they would be identical in the cold plasma limit). Applying these formulae gives $B_0\approx3.5$ kG and 1.5 kG for the kink and the Alfv\'{e}n waves, respectively. The propagation speed of the kink waves resulted in an unrealistically large field strength, which suggests that either the true phase speed is lower than the value determined here, or that the wave is not a pure kink-mode wave, so that Equation~(\ref{equ:B_kink}) is not valid.
We should, however, note that the Bifrost MHD simulations \citep{Gudiksen2011,Carlsson2016} indicate an order of magnitude lower gas densities, averaged over the atmosphere within the 500~km from the solar surface at optical depth unity (i.e., the atmosphere between the two sampled heights). Such diminished densities result in $B_0\approx1.1$ kG and 0.5 kG from the observed speeds of the kink and the Alfv\'{e}n waves, respectively.

We note that the above estimations of the propagation speeds and of the field strengths are based on the time lags determined from the peaks of the phase diagrams. Patches with weaker significance in Figure~\ref{fig:2dstat} would result in much larger phase speeds (particularly for the intensity oscillations) of up to 200~km\,s$^{-1}$. The latter velocities would be too large to be interpreted as a true propagation speed of MHD waves in a magnetic element (they would result in a much too large field strength if approximated with, e.g., Equations~(\ref{equ:B_kink}) or (\ref{equ:B_alfven})). These seemingly very high phase speeds are likely statistically insignificant, and need not concern us too much here. 

In the next section, we review a few relevant theoretical and numerical studies, based on which we will discuss our interpretations of, and speculations on, the observed travel times.

\vspace{3mm}
\section{Comparisons and Discussion}
\label{subsec-waves-comparison}

Theoretical models and numerical simulations of magneto-acoustic and kink waves in photospheric magnetic flux concentrations may be a better way for the interpretation of the observed propagating waves than the simple estimates that we have made in the previous section. However, we should keep in mind that such theoretical investigations have often been confined to a simplified atmosphere whose characteristics may differ, to some extent, from the actual solar atmosphere. We review a few relevant models/simulations in the following that may provide a better understanding of our observations of the propagating, high-frequency, fast transverse and longitudinal waves in small magnetic elements from {\sc Sunrise}/SuFI. We note that the primary drivers of perturbations in the following papers are often motivated by observations of motions as well as the brightness of magnetic bright points.

\textit{Wave excitation models by \citet{Hasan2000}}: They modeled the transverse velocity of magnetic flux tubes excited by their footpoint motion by analytically solving the Klein-Gordon equation for kink waves. They showed that the excitation of kink waves, due to short-duration pulses at the base of the flux tubes, can produce intermittent chromospheric emissions. Such kink waves have been shown to potentially carry enough energy to contribute to coronal heating~\citep[e.g.,][]{Choudhuri1993b,Jafarzadeh2013a}. \citet{Hasan2000} found that these transverse waves cannot represent a major contribution to chromospheric heating unless they are excited by high-frequency motions (periods of $5-50$~s). They speculated that such high-frequency motions could be due to intergranular turbulence (below the photospheric base) that would not be observable from the ground due to the influence of seeing.

The short periods of $43-74$~s that we observed in horizontal-velocity perturbations overlap with the range of periods found necessary by \citet{Hasan2000} for the transverse (kink) waves to transport enough energy to heat the chromosphere.

\textit{Numerical simulations by \citet{Bogdan2003}}: As a continuation of the work by \citet{Rosenthal2002}, \citet{Bogdan2003} extensively studied the propagation of magneto-acoustic-gravity waves within the photospheric and low chromospheric regions of a 2D magnetized atmosphere. In their numerical simulations, uniform waves are generated at a source in the photosphere, that is confined to a $400$~km region within the flux element ($180$~km below the surface), with a driving frequency of $42$~mHz (a period of $23.8$~s). The propagation of horizontal and of vertical perturbations were studied separately. In addition, they investigated wave propagation in magnetic elements with different field strengths, such that the $\beta=1$ layer was located below or above the wave sources. They concluded that the slow and fast waves are decoupled in the low- and high-$\beta$ parts of the atmosphere but that they couple, leading to mode conversion and/or mode transmission, in areas with $\beta=1$, i.e., where the gas and magnetic pressures are comparable. They also discussed that the intermediate Alfv\'{e}n waves, that propagate in both low- and high-$\beta$ regions, may remain coupled with the fast mode through the whole $\beta<1$ domains. They approximated the structure of the magnetic field by a potential field, thus ignoring the current sheet that separates magnetic flux concentrations from their surroundings.

\textit{MHD models by \citet{Hasan2008}}: As a continuation of earlier works by \citet{Hasan2003} and \citet{Hasan2005}, \citet{Hasan2008} proposed that the intensity enhancement of Ca~{\sc ii}~H MBPs are due to a heating process caused by the dissipation of high-frequency, slow magneto-acoustic waves (i.e., frequencies $>10$~mHz) launched as kink waves at the base of the magnetic flux tubes. Propagation of both slow and fast waves were indeed observed in their simulations. They explained the generation of both types of waves in terms of mode conversion and mode transmission at the $\beta=1$ level where (1) horizontal (transverse) motions at the base of the photospheric flux tubes (producing slow, MHD, kink waves in $\beta>1$ regions) are partially converted to slow (longitudinal) acoustic waves propagating along magnetic field lines in $\beta<1$ regions, (2) slow, MHD, transverse waves in the $\beta>1$ medium are partially transmitted as transverse, MHD, fast waves in the $\beta<1$ regions, and (3) acoustic waves generated in the ambient medium are converted into transverse fast modes. \citet{Hasan2008} showed that while the fast transverse waves rapidly travel to the upper layers of the atmosphere, the slow, compressive acoustic waves form shocks at chromospheric heights accompanied by large temperature perturbations (around $900~K$) representing the intensity enhancement of the Ca~{\sc ii}~H MBPs. We conjecture that the high-frequency kink as well as the longitudinal waves we observed here (or at least a fraction of them) are a result of such mode conversion and/or mode transmission processes explained by, e.g., \citet{Bogdan2003} and \citet{Hasan2008}.

\textit{Numerical simulations by \citet{Nutto2012}}: they investigated the unexpectedly small time lags measured between two photospheric layers, similar to those observed by, e.g., \citet{Finsterle2004}. They consider a dynamic atmosphere (i.e., convectively unstable, time-dependent and magnetic) in their numerical simulations and investigate the propagation of both spherical and plane-parallel waves, excited from different locations with respect to small magnetic elements. They found that when one of the height levels (in the solar photosphere at which the waves are observed), or even both of them, are above the $\beta=1$ level, the wave travel time significantly decreases. They explain such a decrease in time lag as the effects of: (1) convergence of formation heights inside the strong magnetic elements, (2) conversion to and detection of the fast magneto-acoustic wave above the $\beta=1$ level, and (3) the refractive propagation path of the fast magneto-acoustic wave, which can lead to very high phase speeds. The refractive propagation is produced in the presence of the inhomogeneous magnetic field, typically associated with magnetic elements. Thus a fast magneto-acoustic wave that starts out propagating upward inside a part of a flux tube, will with time and height turn into a more horizontally directed propagation path. In this case, the phase speeds would be miscalculated since the observed travel time corresponds to an obliquely propagating wave with wave fronts straddling both the upper and the lower height levels.

\citet{Nutto2012} show that the longitudinal oscillations include both slow acoustic and fast magnetic modes in the $\beta<1$ region. Unlike \citet{Hasan2008}, who primarily launched the wave by displacing the base of the flux tube in their MHD model, the wave source in the simulations of \citet{Nutto2012} only has an acoustic nature. Therefore, the incident acoustic wave at the $\beta=1$ level, is partially converted to the fast magneto-acoustic wave and is partially transmitted as the slow acoustic wave propagating to the $\beta<1$ regions. \citet{Nutto2012} also noted that since the dynamic and complex $\tau=1$ level in the actual solar atmosphere may cause several $\beta=1$ levels above each other, several mode conversions may occur.

With respect to the quiet-Sun, the formation height inside magnetic elements of the 300~nm bandpass is expected to decrease due to spectral line weakening in magnetic elements, whereas it is expected to increase due to the enhanced brightness of the Ca~{\sc ii}~H line core in magnetic elements, which strengthens the line emission~\citep{Jafarzadeh2013a}. Hence, we expect the formation height to diverge rather than to converge when going from quiet-Sun to a flux-tube atmosphere. The refractive wave path could be responsible for the seemingly short travel times and the resulting determined high phase speeds of the fast mode in our MBPs.

Summarizing, simulations have shown that (mainly) two types of wave modes (i.e., fast and slow) are generated at the base of small-scale magnetic flux tubes due to impacts from the side or lateral shaking while moving within intergranular lanes~\citep{Bogdan2003}. Bursts of perturbations can be caused by, e.g., expansion or explosion of neighboring granules as well as by intergranular turbulence, with a large range of speeds. The turbulence in the intergranular areas has been speculated to be responsible for the high-frequency range of the generated waves~\citep{Hasan2000}. Pulse-like events (sometimes super-sonic; \citealt{Jafarzadeh2013a}) can excite transverse kink waves traveling along the flux tubes. In addition, the interaction of the magnetic elements with the convective flows generates magneto-acoustic waves that propagate along the field lines.

The following waves can propagate in the $\beta<1$ region: (1) slow, longitudinal, acoustic waves, (2) fast, transverse, magnetic (kink) waves, and (3) fast, longitudinal magnetic waves. Therefore, we may interpret our observed propagating high-frequency (fast) transverse and (fast) longitudinal waves as a result of mode conversions/transmissions.

When the magnetic field is slightly inclined, fast waves can be converted to the Alfv\'{e}n mode in layers well above the $\beta=1$ level, i.e., at the fast wave reflection point~\citep{Cally2011,Khomenko2012}. The MBPs under study here do have small magnetic field inclinations with respect to the line of sight (with an average value of $9\pm4^{\circ}$), similar to those studied by~\citet{Jafarzadeh2014b}. If the reflection point of fast waves reaches the transition region, up to $30\%$ of the fast wave's energy flux can be carried across the transition region by the Alfv\'{e}n waves due to the fast-to-Alfv\'{e}n mode conversion~\citep{Hansen2012}. Thus, fast waves may contribute to the heating of the outer atmospheric layers. We cannot, however, verify whether the fast waves we observed in the magnetic elements can reach these atmospheric heights since our data do not sample those layers.

We note that this interpretation should be treated with caution. \citet{Bogdan2003} explained characteristics of oscillations in the solar magneto-atmosphere using extensive numerical simulations. They showed that at any given location, a superposition of several distinct waves with different characteristics may be observed. These waves may propagate in different directions and may come from different locations, e.g., directly from their sources, from the equipartition level as a result of mode conversion, or locally from the interaction of various kinds of waves with $p$-modes. Therefore, distinguishing the nature of different waves may not be straightforward from observations alone.

Similar to our findings, \citet{Morton2012} have also observed both transverse (incompressible) and longitudinal (compressible) waves along magnetic flux tubes close to the quiet-Sun disk-center, sampled in the mid-chromosphere. They interpreted these two types of waves as ($1$) the fast MHD kink mode, measured from transverse displacements of the magnetic structures, and ($2$) the fast MHD sausage mode, determined from intensity perturbations. Our observations may represent the lower solar atmosphere's origin of the fast waves observed in the mid to upper chromosphere by \citet{Morton2012}. They found propagation speeds of $\approx60-90$~km\,s$^{-1}$ for the kink oscillations and a range of $\approx50-320$~km\,s$^{-1}$ for the longitudinal oscillations. The typical periods of $180-210$~s for the fast kink waves and of $90-190$~s for the fast longitudinal waves measured by \citet{Morton2012} are somewhat larger than the ones we obtain here (with an overlap for the longitudinal oscillations). The size variations of our point-like magnetic elements are too small, to investigate the relationship between size and intensity variations to detect any manifestation of the sausage mode in the intensity perturbations. For somewhat larger magnetic features observed by {\sc Sunrise}, however, \citet{Martinez-Gonzalez2011} have noticed such oscillations.

The upward wave propagation in small MBPs, studied here, agrees with that of \citet{Jess2012} who found a similar behavior in intensity oscillations of a larger number of MBPs in both observations and simulations. They investigated the wave propagation in all pixels of a relatively large FOV and concluded that much of the power is concentrated in MBPs. However, with the lower spatial resolution of their observations compared to {\sc Sunrise}/SuFI, they found only low-frequency waves in the rage of $1.7-7$~mHz in intensity oscillations.
The spatial resolution has been shown to have a direct correlation with the power of high-frequency oscillations, particularly on small spatial scales, with lower spatial resolution leading to a decreased sensitivity to power \citep{Wedemeyer-Bohm2007}.
The factor of two longer time series of observational data used by \citet{Jess2012} has consequently allowed them to obtain such low frequencies. However, \citet{Jess2012} also found low-frequency intensity oscillations on the order of $2-9$~mHz from their MHD simulations with grid size of 25~km (i.e., spatial resolution of 50~km) and a similar frequency resolution as in our study.

\citet{Jafarzadeh2017a} and \citet{Gafeira2017b} studied transverse oscillations and sausage-mode waves in bright, slender fibrils observed in a narrowband Ca~{\sc ii}~H passband (FWHM$\approx0.11$~nm) provided by SuFI during the second flight of the {\sc Sunrise} observatory \citep{Solanki2017}. They found periods in the range of $\approx20-160$~s for both types of waves, with median values of roughly 83~s and 34~s for the transverse and sausage waves, respectively. Their range of periods overlaps with those of the MBPs we observe in this study. The slender Ca~{\sc ii}~H fibrils have been shown to outline the nearly horizontal magnetic field lines in the low solar chromosphere, slightly higher than the heights, where our Ca~{\sc ii}~H MBPs were sampled \citep{Jafarzadeh2017b}. Since such fibrils have footpoints in photospheric magnetic features, such as MBPs, we speculate that waves of the type observed here partly continue into the chromosphere and may become visible as oscillations of fibrils.

\vspace{4mm}
\section{Conclusions}
\label{sec-waves-conclusions}

We studied oscillations in the horizontal displacement and intensity of small-scale MBPs as observed in image sequences taken in the passbands at 300~nm and in the Ca~{\sc ii}~H spectral line with {\sc Sunrise}/SuFI. Time-series of seven MBPs reliably displaying oscillations of both types were analyzed. Although such a small number of MBPs may not be representative of quiet-Sun magnetic elements in general, the obtained properties of the oscillatory motions are free of effects introduced by seeing.

We computed a height difference of $450\pm100$~km between the two atmospheric layers sampled by the 300~nm and Ca~{\sc ii}~H passbands by employing the RH radiative transfer code of \citet{Uitenbroek2001}. A comparable height difference was obtained from the analysis of the intensity oscillations, likely due to acoustic waves, in a $14\arcsec\times5\arcsec$ quiet-Sun area. The phase diagram of the propagating high-frequency (4-38~mHz) acoustic waves resulted in a time lag of $57\pm2$~s, implying an average height difference of $430\pm30$~km between the two atmospheric layers sampled by the {\sc Sunrise}/SuFI 300~nm and Ca~{\sc ii}~H passbands.

Our wavelet analysis of the small-scale magnetic elements yields: ($1$) consistent oscillations with high frequencies of up to $30$~mHz in intensity and up to $23$~mHz in horizontal displacement (not limited by the Nyquist frequency of $\approx42$~mHz), ($2$) positive phase-lags between both kinds of oscillations, i.e., upward propagation of the waves, ($3$) fast waves with a phase speed of $29\pm2$~km\,s$^{-1}$ in the horizontal displacements (kink mode) and a propagating speed of $31\pm2$~km\,s$^{-1}$ in intensity oscillations (longitudinal waves), and ($4$) a relatively wide range of phase spectra, which may describe different dispersion relations and/or belong to different sources~\citep{Bogdan2003}. In addition, there is a slight indication of standing and downward propagating waves in the intensity oscillations.

Fast waves are of interest because they can propagate to the upper solar atmosphere and carry energy. By comparing our results with those from theoretical investigations (see Section~\ref{subsec-waves-comparison}), observations of the fast waves could possibly be explained as a result of mode conversion and/or mode transmission at the $\beta=1$ level (i.e., where Alfv\'{e}n and sound speeds coincide;~\citealt{Bogdan2003}). 
The observed high-frequency waves in the magnetic elements could be excited due to, e.g., buffeting of flux-tube footpoints by high-frequency perturbations caused by the surrounding granules and by intergranular turbulence~\citep{Hasan2008}.

Summarizing, we have observed high-frequency, fast, upward propagating waves using data from {\sc Sunrise} unaffected by seeing. 
Fast waves in both, horizontal displacement and intensity oscillations appear to dominate over slow waves in our data. We speculate that the rather large propagation speeds that we deduced from our data (particularly those obtained from smaller concentrations in the phase diagram of Figure~\ref{fig:2dstat}) could also be due to (1) refraction of the propagation path above the magnetic canopy for longitudinal waves~\citep{Nutto2012}, and (2) to the superposition of several wave trains, e.g., polarized kink waves for transverse waves~\citep{Bogdan2003,Fujimura2009}. 

We cannot completely rule out that we have overestimated the formation height of the Ca~{\sc ii}~H passband, which may have led to overly large phase speeds. 
This is because the plane-parallel atmospheres used to compute the contribution functions may not be representative of the complex 3D solar atmosphere. Consequently, we determined the height difference between the two wavelength bands by two entirely independent means. Besides the contribution functions, we also used the time difference between the arrival of acoustic waves in the quiet-Sun at the two heights. Both methods gave very similar results in the quiet-Sun, which greatly increases our confidence in our conclusions.
In order to find out more about the actual cause of these fast waves, a similar study as was done here, but using synthetic passbands from 3D MHD simulation is essential. Further studies from, e.g., simultaneous observations of multiple atmospheric layers (from the photosphere to the transitions-region/corona) might clarify to what extent the high-frequency fast waves can reach the upper solar atmosphere and how much energy they release at those heights.

\acknowledgements
The German contribution to {\sc Sunrise} is funded by the Bundesministerium f\"{u}r Wirtschaft und Technologie through the Deutsches Zentrum f\"{u}r Luft- und Raumfahrt e.V. (DLR), grant No. 50 OU 0401, and by the Innovationsfond of the President of the Max Planck Society (MPG). The Spanish contribution has been funded by the Spanish MICINN under projects ESP2006-13030-C06 and AYA2009-14105-C06 (including European FEDER funds). The HAO contribution was partly funded through NASA grant NNX08AH38G. This work was partly supported by the BK21 plus program through the National Research Foundation (NRF) funded by the Ministry of Education of Korea. S.J. receives support from the Research Council of Norway.

\vspace{3mm}
\bibliographystyle{aa}
\bibliography{Shahin}

\begin{thebibliography}{115}
\expandafter\ifx\csname natexlab\endcsname\relax\def\natexlab#1{#1}\fi

\bibitem[{{Anusha} {et~al.}(2017){Anusha}, {Solanki}, {Hirzberger}, \&
  {Feller}}]{Anusha2017}
{Anusha}, L.~S., {Solanki}, S.~K., {Hirzberger}, J., \& {Feller}, A. 2017,
  \aap, 598, A47

\bibitem[{{Aschwanden} {et~al.}(1999){Aschwanden}, {Fletcher}, {Schrijver}, \&
  {Alexander}}]{Aschwanden1999}
{Aschwanden}, M.~J., {Fletcher}, L., {Schrijver}, C.~J., \& {Alexander}, D.
  1999, \apj, 520, 880

\bibitem[{{Ayres} {et~al.}(1986){Ayres}, {Testerman}, \& {Brault}}]{Ayres1986}
{Ayres}, T.~R., {Testerman}, L., \& {Brault}, J.~W. 1986, \apj, 304, 542

\bibitem[{{Barthol} {et~al.}(2011){Barthol}, {Gandorfer}, {Solanki},
  {Sch{\"u}ssler}, {Chares}, {Curdt}, {Deutsch}, {Feller}, {Germerott},
  {Grauf}, {Heerlein}, {Hirzberger}, {Kolleck}, {Meller}, {M{\"u}ller},
  {Riethm{\"u}ller}, {Tomasch}, {Kn{\"o}lker}, {Lites}, {Card}, {Elmore},
  {Fox}, {Lecinski}, {Nelson}, {Summers}, {Watt}, {Mart{\'{\i}}nez Pillet},
  {Bonet}, {Schmidt}, {Berkefeld}, {Title}, {Domingo}, {Gasent Blesa}, {Del
  Toro Iniesta}, {L{\'o}pez Jim{\'e}nez}, {{\'A}lvarez-Herrero},
  {Sabau-Graziati}, {Widani}, {Haberler}, {H{\"a}rtel}, {Kampf}, {Levin},
  {P{\'e}rez Grande}, {Sanz-Andr{\'e}s}, \& {Schmidt}}]{Barthol2011}
{Barthol}, P., {Gandorfer}, A., {Solanki}, S.~K., {et~al.} 2011, \solphys, 268,
  1

\bibitem[{{Baudin} {et~al.}(1996){Baudin}, {Bocchialini}, \&
  {Koutchmy}}]{Baudin1996}
{Baudin}, F., {Bocchialini}, K., \& {Koutchmy}, S. 1996, \aap, 314, L9

\bibitem[{{Baudin} {et~al.}(1994){Baudin}, {Gabriel}, \& {Gibert}}]{Baudin1994}
{Baudin}, F., {Gabriel}, A., \& {Gibert}, D. 1994, \aap, 285, L29

\bibitem[{{Beck} {et~al.}(2008){Beck}, {Schmidt}, {Rezaei}, \&
  {Rammacher}}]{Beck2008}
{Beck}, C., {Schmidt}, W., {Rezaei}, R., \& {Rammacher}, W. 2008, \aap, 479,
  213

\bibitem[{{Bel} \& {Leroy}(1977)}]{Bel1977}
{Bel}, N. \& {Leroy}, B. 1977, \aap, 55, 239

\bibitem[{{Bello Gonz{\'a}lez} {et~al.}(2009){Bello Gonz{\'a}lez}, {Flores
  Soriano}, {Kneer}, \& {Okunev}}]{BelloGonzalez2009}
{Bello Gonz{\'a}lez}, N., {Flores Soriano}, M., {Kneer}, F., \& {Okunev}, O.
  2009, \aap, 508, 941

\bibitem[{{Bello Gonz{\'a}lez} {et~al.}(2010){Bello Gonz{\'a}lez}, {Franz},
  {Mart{\'{\i}}nez Pillet}, {Bonet}, {Solanki}, {del Toro Iniesta}, {Schmidt},
  {Gandorfer}, {Domingo}, {Barthol}, {Berkefeld}, \&
  {Kn{\"o}lker}}]{BelloGonzalez2010}
{Bello Gonz{\'a}lez}, N., {Franz}, M., {Mart{\'{\i}}nez Pillet}, V., {et~al.}
  2010, \apjl, 723, L134

\bibitem[{{Berkefeld} {et~al.}(2011){Berkefeld}, {Schmidt}, {Soltau}, {Bell},
  {Doerr}, {Feger}, {Friedlein}, {Gerber}, {Heidecke}, {Kentischer},
  {von~der~L{\"u}he}, {Sigwarth}, {W{\"a}lde}, {Barthol}, {Deutsch},
  {Gandorfer}, {Germerott}, {Grauf}, {Meller}, {{\'A}lvarez-Herrero},
  {Kn{\"o}lker}, {Mart{\'{\i}}nez Pillet}, {Solanki}, \&
  {Title}}]{Berkefeld2011}
{Berkefeld}, T., {Schmidt}, W., {Soltau}, D., {et~al.} 2011, \solphys, 268, 103

\bibitem[{{Bloomfield} {et~al.}(2004){Bloomfield}, {McAteer}, {Mathioudakis},
  {Williams}, \& {Keenan}}]{Bloomfield2004}
{Bloomfield}, D.~S., {McAteer}, R.~T.~J., {Mathioudakis}, M., {Williams},
  D.~R., \& {Keenan}, F.~P. 2004, \apj, 604, 936

\bibitem[{{Bogdan} {et~al.}(2003){Bogdan}, {Carlsson}, {Hansteen}, {McMurry},
  {Rosenthal}, {Johnson}, {Petty-Powell}, {Zita}, {Stein}, {McIntosh}, \&
  {Nordlund}}]{Bogdan2003}
{Bogdan}, T.~J., {Carlsson}, M., {Hansteen}, V.~H., {et~al.} 2003, \apj, 599,
  626

\bibitem[{{Cally}(2007)}]{Cally2007}
{Cally}, P.~S. 2007, Astronomische Nachrichten, 328, 286

\bibitem[{{Cally} \& {Hansen}(2011)}]{Cally2011}
{Cally}, P.~S. \& {Hansen}, S.~C. 2011, \apj, 738, 119

\bibitem[{{Carlsson} {et~al.}(2016){Carlsson}, {Hansteen}, {Gudiksen},
  {Leenaarts}, \& {De Pontieu}}]{Carlsson2016}
{Carlsson}, M., {Hansteen}, V.~H., {Gudiksen}, B.~V., {Leenaarts}, J., \& {De
  Pontieu}, B. 2016, \aap, 585, A4

\bibitem[{{Carlsson} \& {Stein}(1997)}]{Carlsson1997}
{Carlsson}, M. \& {Stein}, R.~F. 1997, \apj, 481, 500

\bibitem[{{Centeno} {et~al.}(2006){Centeno}, {Collados}, \& {Trujillo
  Bueno}}]{Centeno2006}
{Centeno}, R., {Collados}, M., \& {Trujillo Bueno}, J. 2006, \apj, 640, 1153

\bibitem[{{Choudhuri} {et~al.}(1993{\natexlab{a}}){Choudhuri}, {Auffret}, \&
  {Priest}}]{Choudhuri1993a}
{Choudhuri}, A.~R., {Auffret}, H., \& {Priest}, E.~R. 1993{\natexlab{a}},
  \solphys, 143, 49

\bibitem[{{Choudhuri} {et~al.}(1993{\natexlab{b}}){Choudhuri}, {Dikpati}, \&
  {Banerjee}}]{Choudhuri1993b}
{Choudhuri}, A.~R., {Dikpati}, M., \& {Banerjee}, D. 1993{\natexlab{b}}, \apj,
  413, 811

\bibitem[{{Cranmer} \& {van Ballegooijen}(2005)}]{Cranmer2005}
{Cranmer}, S.~R. \& {van Ballegooijen}, A.~A. 2005, \apjs, 156, 265

\bibitem[{{Daubechies}(1990)}]{Daubechies1990}
{Daubechies}, I. 1990, IEEE Trans. Inf. Theory, 36, 961

\bibitem[{{De Moortel} {et~al.}(2002){De Moortel}, {Ireland}, {Hood}, \&
  {Walsh}}]{DeMoortel2002}
{De Moortel}, I., {Ireland}, J., {Hood}, A.~W., \& {Walsh}, R.~W. 2002, \aap,
  387, L13

\bibitem[{{De Pontieu} {et~al.}(2004){De Pontieu}, {Erd{\'e}lyi}, \&
  {James}}]{DePontieu2004a}
{De Pontieu}, B., {Erd{\'e}lyi}, R., \& {James}, S.~P. 2004, \nat, 430, 536

\bibitem[{{De Pontieu} {et~al.}(2007){De Pontieu}, {McIntosh}, {Carlsson},
  {Hansteen}, {Tarbell}, {Schrijver}, {Title}, {Shine}, {Tsuneta}, {Katsukawa},
  {Ichimoto}, {Suematsu}, {Shimizu}, \& {Nagata}}]{DePontieu2007}
{De Pontieu}, B., {McIntosh}, S.~W., {Carlsson}, M., {et~al.} 2007, Science,
  318, 1574

\bibitem[{{de Wijn} {et~al.}(2005){de Wijn}, {Rutten}, {Haverkamp}, \&
  {S{\"u}tterlin}}]{de-Wijn2005}
{de Wijn}, A.~G., {Rutten}, R.~J., {Haverkamp}, E.~M.~W.~P., \&
  {S{\"u}tterlin}, P. 2005, \aap, 441, 1183

\bibitem[{{DeForest}(2004)}]{DeForest2004}
{DeForest}, C.~E. 2004, \apjl, 617, L89

\bibitem[{{Deubner} \& {Gough}(1984)}]{Deubner1984}
{Deubner}, F.-L. \& {Gough}, D. 1984, \araa, 22, 593

\bibitem[{{Dorotovi{\v c}} {et~al.}(2014){Dorotovi{\v c}}, {Erd{\'e}lyi},
  {Freij}, {Karlovsk{\'y}}, \& {M{\'a}rquez}}]{Dorotovic2014}
{Dorotovi{\v c}}, I., {Erd{\'e}lyi}, R., {Freij}, N., {Karlovsk{\'y}}, V., \&
  {M{\'a}rquez}, I. 2014, \aap, 563, A12

\bibitem[{{Edwin} \& {Roberts}(1983)}]{Edwin1983}
{Edwin}, P.~M. \& {Roberts}, B. 1983, \solphys, 88, 179

\bibitem[{{Erd{\'e}lyi} {et~al.}(2007){Erd{\'e}lyi}, {Malins}, {T{\'o}th}, \&
  {de Pontieu}}]{Erdelyi2007}
{Erd{\'e}lyi}, R., {Malins}, C., {T{\'o}th}, G., \& {de Pontieu}, B. 2007,
  \aap, 467, 1299

\bibitem[{{Farge}(1992)}]{Farge1992}
{Farge}, M. 1992, Ann. Rev. Fluid Mech., 24, 395

\bibitem[{{Fedun} {et~al.}(2011){Fedun}, {Shelyag}, \&
  {Erd{\'e}lyi}}]{Fedun2011}
{Fedun}, V., {Shelyag}, S., \& {Erd{\'e}lyi}, R. 2011, \apj, 727, 17

\bibitem[{{Felipe}(2012)}]{Felipe2012}
{Felipe}, T. 2012, \apj, 758, 96

\bibitem[{{Finsterle} {et~al.}(2004){Finsterle}, {Jefferies}, {Cacciani}, \&
  {Rapex}}]{Finsterle2004}
{Finsterle}, W., {Jefferies}, S.~M., {Cacciani}, A., \& {Rapex}, P. 2004, in
  ESA Special Pub. (New Haven: ESA SP), Vol. 559, SOHO 14 Helio- and
  Asteroseismology: Towards a Golden Future, ed. {{Danesy}, D.}, 223

\bibitem[{{Fontenla} {et~al.}(1993){Fontenla}, {Avrett}, \&
  {Loeser}}]{Fontenla1993}
{Fontenla}, J.~M., {Avrett}, E.~H., \& {Loeser}, R. 1993, \apj, 406, 319

\bibitem[{{Fossum} \& {Carlsson}(2005)}]{Fossum2005}
{Fossum}, A. \& {Carlsson}, M. 2005, \nat, 435, 919

\bibitem[{{Fujimura} \& {Tsuneta}(2009)}]{Fujimura2009}
{Fujimura}, D. \& {Tsuneta}, S. 2009, \apj, 702, 1443

\bibitem[{{Gafeira} {et~al.}(2017){Gafeira}, {Jafarzadeh}, {Solanki}, {Lagg},
  {van Noort}, {Barthol}, \& {Sunrise team}}]{Gafeira2017b}
{Gafeira}, R., {Jafarzadeh}, S., {Solanki}, S.~K., {et~al.} 2017, \apjs, 229, 7

\bibitem[{{Gandorfer} {et~al.}(2011){Gandorfer}, {Grauf}, {Barthol},
  {Riethm{\"u}ller}, {Solanki}, {Chares}, {Deutsch}, {Ebert}, {Feller},
  {Germerott}, {Heerlein}, {Heinrichs}, {Hirche}, {Hirzberger}, {Kolleck},
  {Meller}, {M{\"u}ller}, {Sch{\"a}fer}, {Tomasch}, {Kn{\"o}lker},
  {Mart{\'{\i}}nez Pillet}, {Bonet}, {Schmidt}, {Berkefeld}, {Feger},
  {Heidecke}, {Soltau}, {Tischenberg}, {Fischer}, {Title}, {Anwand}, \&
  {Schmidt}}]{Gandorfer2011}
{Gandorfer}, A., {Grauf}, B., {Barthol}, P., {et~al.} 2011, \solphys, 268, 35

\bibitem[{{Grinsted} {et~al.}(2004){Grinsted}, {Moore}, \&
  {Jevrejeva}}]{Grinsted2004}
{Grinsted}, A., {Moore}, J.~C., \& {Jevrejeva}, S. 2004, Nonlinear Process.
  Geophys., 11, 561

\bibitem[{{Grossmann-Doerth} {et~al.}(1988){Grossmann-Doerth}, {Larsson}, \&
  {Solanki}}]{GrossmannDoerth1988}
{Grossmann-Doerth}, U., {Larsson}, B., \& {Solanki}, S.~K. 1988, \aap, 204, 266

\bibitem[{{Grubbs}(1969)}]{Grubbs1969}
{Grubbs}, F. 1969, Technometrics, 11, 1

\bibitem[{{Gudiksen} {et~al.}(2011){Gudiksen}, {Carlsson}, {Hansteen}, {Hayek},
  {Leenaarts}, \& {Mart{\'{\i}}nez-Sykora}}]{Gudiksen2011}
{Gudiksen}, B.~V., {Carlsson}, M., {Hansteen}, V.~H., {et~al.} 2011, \aap, 531,
  A154

\bibitem[{{Hansen} \& {Cally}(2009)}]{Hansen2009}
{Hansen}, S.~C. \& {Cally}, P.~S. 2009, \solphys, 255, 193

\bibitem[{{Hansen} \& {Cally}(2012)}]{Hansen2012}
{Hansen}, S.~C. \& {Cally}, P.~S. 2012, \apj, 751, 31

\bibitem[{{Hasan} \& {Kalkofen}(1999)}]{Hasan1999}
{Hasan}, S.~S. \& {Kalkofen}, W. 1999, \apj, 519, 899

\bibitem[{{Hasan} {et~al.}(2000){Hasan}, {Kalkofen}, \& {van
  Ballegooijen}}]{Hasan2000}
{Hasan}, S.~S., {Kalkofen}, W., \& {van Ballegooijen}, A.~A. 2000, \apjl, 535,
  L67

\bibitem[{{Hasan} {et~al.}(2003){Hasan}, {Kalkofen}, {van Ballegooijen}, \&
  {Ulmschneider}}]{Hasan2003}
{Hasan}, S.~S., {Kalkofen}, W., {van Ballegooijen}, A.~A., \& {Ulmschneider},
  P. 2003, \apj, 585, 1138

\bibitem[{{Hasan} \& {Sobouti}(1987)}]{Hasan1987}
{Hasan}, S.~S. \& {Sobouti}, Y. 1987, \mnras, 228, 427

\bibitem[{{Hasan} \& {van Ballegooijen}(2008)}]{Hasan2008}
{Hasan}, S.~S. \& {van Ballegooijen}, A.~A. 2008, \apj, 680, 1542

\bibitem[{{Hasan} {et~al.}(2005){Hasan}, {van Ballegooijen}, {Kalkofen}, \&
  {Steiner}}]{Hasan2005}
{Hasan}, S.~S., {van Ballegooijen}, A.~A., {Kalkofen}, W., \& {Steiner}, O.
  2005, \apj, 631, 1270

\bibitem[{{He} {et~al.}(2009){He}, {Marsch}, {Tu}, \& {Tian}}]{He2009}
{He}, J., {Marsch}, E., {Tu}, C., \& {Tian}, H. 2009, \apjl, 705, L217

\bibitem[{{Hirzberger} {et~al.}(2011){Hirzberger}, {Feller}, {Riethm{\"u}ller},
  {Gandorfer}, \& {Solanki}}]{Hirzberger2011}
{Hirzberger}, J., {Feller}, A., {Riethm{\"u}ller}, T.~L., {Gandorfer}, A., \&
  {Solanki}, S.~K. 2011, \aap, 529, A132

\bibitem[{{Hirzberger} {et~al.}(2010){Hirzberger}, {Feller}, {Riethm{\"u}ller},
  {Sch{\"u}ssler}, {Borrero}, {Afram}, {Unruh}, {Berdyugina}, {Gandorfer},
  {Solanki}, {Barthol}, {Bonet}, {Mart{\'{\i}}nez Pillet}, {Berkefeld},
  {Kn{\"o}lker}, {Schmidt}, \& {Title}}]{Hirzberger2010}
{Hirzberger}, J., {Feller}, A., {Riethm{\"u}ller}, T.~L., {et~al.} 2010, \apjl,
  723, L154

\bibitem[{{Holzreuter} \& {Solanki}(2015)}]{Holzreuter2015}
{Holzreuter}, R. \& {Solanki}, S.~K. 2015, \aap, 582, A101

\bibitem[{{Jafarzadeh} {et~al.}(2017{\natexlab{a}}){Jafarzadeh}, {Rutten},
  {Solanki}, {Wiegelmann}, {Riethm{\"u}ller}, {van Noort}, {Szydlarski},
  {Blanco Rodr{\'\i}guez}, {Barthol}, {del Toro Iniesta}, \& {Sunrise
  team}}]{Jafarzadeh2017b}
{Jafarzadeh}, S., {Rutten}, R.~J., {Solanki}, S.~K., {et~al.}
  2017{\natexlab{a}}, \apjs, 229, 11

\bibitem[{{Jafarzadeh} {et~al.}(2013){Jafarzadeh}, {Solanki}, {Feller}, {Lagg},
  {Pietarila}, {Danilovic}, {Riethm{\"u}ller}, \& {Mart{\'{\i}}nez
  Pillet}}]{Jafarzadeh2013a}
{Jafarzadeh}, S., {Solanki}, S.~K., {Feller}, A., {et~al.} 2013, \aap, 549,
  A116

\bibitem[{{Jafarzadeh} {et~al.}(2017{\natexlab{b}}){Jafarzadeh}, {Solanki},
  {Gafeira}, {van Noort}, {Barthol}, {Berkefeld}, \& {Sunrise
  team}}]{Jafarzadeh2017a}
{Jafarzadeh}, S., {Solanki}, S.~K., {Gafeira}, R., {et~al.} 2017{\natexlab{b}},
  \apjs, 229, 9

\bibitem[{{Jafarzadeh} {et~al.}(2014){Jafarzadeh}, {Solanki}, {Lagg}, {Bellot
  Rubio}, {van Noort}, {Feller}, \& {Danilovic}}]{Jafarzadeh2014b}
{Jafarzadeh}, S., {Solanki}, S.~K., {Lagg}, A., {et~al.} 2014, \aap, 569, A105

\bibitem[{{Jefferies} {et~al.}(2006){Jefferies}, {McIntosh}, {Armstrong},
  {Bogdan}, {Cacciani}, \& {Fleck}}]{Jefferies2006}
{Jefferies}, S.~M., {McIntosh}, S.~W., {Armstrong}, J.~D., {et~al.} 2006,
  \apjl, 648, L151

\bibitem[{{Jess} {et~al.}(2007){Jess}, {Andi{\'c}}, {Mathioudakis},
  {Bloomfield}, \& {Keenan}}]{Jess2007}
{Jess}, D.~B., {Andi{\'c}}, A., {Mathioudakis}, M., {Bloomfield}, D.~S., \&
  {Keenan}, F.~P. 2007, \aap, 473, 943

\bibitem[{{Jess} {et~al.}(2009){Jess}, {Mathioudakis}, {Erd{\'e}lyi},
  {Crockett}, {Keenan}, \& {Christian}}]{Jess2009}
{Jess}, D.~B., {Mathioudakis}, M., {Erd{\'e}lyi}, R., {et~al.} 2009, Science,
  323, 1582

\bibitem[{{Jess} {et~al.}(2012){Jess}, {Shelyag}, {Mathioudakis}, {Keys},
  {Christian}, \& {Keenan}}]{Jess2012}
{Jess}, D.~B., {Shelyag}, S., {Mathioudakis}, M., {et~al.} 2012, \apj, 746, 183

\bibitem[{{Kato} {et~al.}(2016){Kato}, {Steiner}, {Hansteen}, {Gudiksen},
  {Wedemeyer}, \& {Carlsson}}]{Kato2016}
{Kato}, Y., {Steiner}, O., {Hansteen}, V., {et~al.} 2016, \apj, 827, 7

\bibitem[{{Kato} {et~al.}(2011){Kato}, {Steiner}, {Steffen}, \&
  {Suematsu}}]{Kato2011}
{Kato}, Y., {Steiner}, O., {Steffen}, M., \& {Suematsu}, Y. 2011, \apjl, 730,
  L24

\bibitem[{{Keys} {et~al.}(2013){Keys}, {Mathioudakis}, {Jess}, {Shelyag},
  {Christian}, \& {Keenan}}]{Keys2013}
{Keys}, P.~H., {Mathioudakis}, M., {Jess}, D.~B., {et~al.} 2013, \mnras, 428,
  3220

\bibitem[{{Khomenko} \& {Cally}(2012)}]{Khomenko2012}
{Khomenko}, E. \& {Cally}, P.~S. 2012, \apj, 746, 68

\bibitem[{{Khomenko} {et~al.}(2008){Khomenko}, {Collados}, \&
  {Felipe}}]{Khomenko2008}
{Khomenko}, E., {Collados}, M., \& {Felipe}, T. 2008, \solphys, 251, 589

\bibitem[{{Lawrence} \& {Cadavid}(2012)}]{Lawrence2012}
{Lawrence}, J.~K. \& {Cadavid}, A.~C. 2012, \solphys, 280, 125

\bibitem[{{Lawrence} {et~al.}(2011){Lawrence}, {Cadavid}, {Christian}, {Jess},
  \& {Mathioudakis}}]{Lawrence2011}
{Lawrence}, J.~K., {Cadavid}, A.~C., {Christian}, D.~J., {Jess}, D.~B., \&
  {Mathioudakis}, M. 2011, \apjl, 743, L24

\bibitem[{{Magain}(1986)}]{Magain1986}
{Magain}, P. 1986, \aap, 163, 135

\bibitem[{{Mart{\'{\i}}nez Gonz{\'a}lez} {et~al.}(2011){Mart{\'{\i}}nez
  Gonz{\'a}lez}, {Asensio Ramos}, {Manso Sainz}, {Khomenko}, {Mart{\'{\i}}nez
  Pillet}, {Solanki}, {L{\'o}pez Ariste}, {Schmidt}, {Barthol}, \&
  {Gandorfer}}]{Martinez-Gonzalez2011}
{Mart{\'{\i}}nez Gonz{\'a}lez}, M.~J., {Asensio Ramos}, A., {Manso Sainz}, R.,
  {et~al.} 2011, \apjl, 730, L37

\bibitem[{{Mart{\'{\i}}nez Pillet} {et~al.}(2011){Mart{\'{\i}}nez Pillet}, {Del
  Toro Iniesta}, {{\'A}lvarez-Herrero}, {Domingo}, {Bonet}, {Gonz{\'a}lez
  Fern{\'a}ndez}, {L{\'o}pez Jim{\'e}nez}, {Pastor}, {Gasent Blesa}, {Mellado},
  {Piqueras}, {Aparicio}, {Balaguer}, {Ballesteros}, {Belenguer}, {Bellot
  Rubio}, {Berkefeld}, {Collados}, {Deutsch}, {Feller}, {Girela}, {Grauf},
  {Heredero}, {Herranz}, {Jer{\'o}nimo}, {Laguna}, {Meller}, {Men{\'e}ndez},
  {Morales}, {Orozco Su{\'a}rez}, {Ramos}, {Reina}, {Ramos},
  {Rodr{\'{\i}}guez}, {S{\'a}nchez}, {Uribe-Patarroyo}, {Barthol}, {Gandorfer},
  {Kn{\"o}lker}, {Schmidt}, {Solanki}, \& {Vargas
  Dom{\'{\i}}nguez}}]{Martinez-Pillet2011}
{Mart{\'{\i}}nez Pillet}, V., {Del Toro Iniesta}, J.~C., {{\'A}lvarez-Herrero},
  A., {et~al.} 2011, \solphys, 268, 57

\bibitem[{{Mathioudakis} {et~al.}(2013){Mathioudakis}, {Jess}, \&
  {Erd{\'e}lyi}}]{Mathioudakis2013}
{Mathioudakis}, M., {Jess}, D.~B., \& {Erd{\'e}lyi}, R. 2013, \ssr, 175, 1

\bibitem[{{McIntosh} {et~al.}(2011){McIntosh}, {De Pontieu}, {Carlsson},
  {Hansteen}, {Boerner}, \& {Goossens}}]{McIntosh2011}
{McIntosh}, S.~W., {De Pontieu}, B., {Carlsson}, M., {et~al.} 2011, \nat, 475,
  477

\bibitem[{{McIntosh} \& {Smillie}(2004)}]{McIntosh2004}
{McIntosh}, S.~W. \& {Smillie}, D.~G. 2004, \apj, 604, 924

\bibitem[{{Michalitsanos}(1973)}]{Michalitsanos1973}
{Michalitsanos}, A.~G. 1973, \solphys, 30, 47

\bibitem[{{Morton} {et~al.}(2011){Morton}, {Erd{\'e}lyi}, {Jess}, \&
  {Mathioudakis}}]{Morton2011}
{Morton}, R.~J., {Erd{\'e}lyi}, R., {Jess}, D.~B., \& {Mathioudakis}, M. 2011,
  \apjl, 729, L18

\bibitem[{{Morton} {et~al.}(2012){Morton}, {Verth}, {Jess}, {Kuridze},
  {Ruderman}, {Mathioudakis}, \& {Erd{\'e}lyi}}]{Morton2012}
{Morton}, R.~J., {Verth}, G., {Jess}, D.~B., {et~al.} 2012, Nature
  Communications, 3, 1315

\bibitem[{{Nakariakov} \& {Verwichte}(2005)}]{Nakariakov2005}
{Nakariakov}, V.~M. \& {Verwichte}, E. 2005, Living Rev. Solar Phys., 2, 3.
  URL: http://www.livingreviews.org/lrsp

\bibitem[{{Nutto} {et~al.}(2012){Nutto}, {Steiner}, {Schaffenberger}, \&
  {Roth}}]{Nutto2012}
{Nutto}, C., {Steiner}, O., {Schaffenberger}, W., \& {Roth}, M. 2012, \aap,
  538, A79

\bibitem[{{Okamoto} \& {De Pontieu}(2011)}]{Okamoto2011}
{Okamoto}, T.~J. \& {De Pontieu}, B. 2011, \apjl, 736, L24

\bibitem[{{Pietarila} {et~al.}(2011){Pietarila}, {Aznar Cuadrado},
  {Hirzberger}, \& {Solanki}}]{Pietarila2011}
{Pietarila}, A., {Aznar Cuadrado}, R., {Hirzberger}, J., \& {Solanki}, S.~K.
  2011, \apj, 739, 92

\bibitem[{{Press} {et~al.}(2007){Press}, {Teukolsky}, {Vetterling}, \&
  {Flannery}}]{Press2007}
{Press}, W.~H., {Teukolsky}, S.~A., {Vetterling}, W.~T., \& {Flannery}, B.~P.
  2007, {Numerical recipes: The art of scientific computing (Cambridge, UK:
  Cambridge University Press)}

\bibitem[{{Riethm{\"u}ller} {et~al.}(2010){Riethm{\"u}ller}, {Solanki},
  {Mart{\'{\i}}nez Pillet}, {Hirzberger}, {Feller}, {Bonet}, {Bello
  Gonz{\'a}lez}, {Franz}, {Sch{\"u}ssler}, {Barthol}, {Berkefeld}, {del Toro
  Iniesta}, {Domingo}, {Gandorfer}, {Kn{\"o}lker}, \&
  {Schmidt}}]{Riethmuller2010}
{Riethm{\"u}ller}, T.~L., {Solanki}, S.~K., {Mart{\'{\i}}nez Pillet}, V.,
  {et~al.} 2010, \apjl, 723, L169

\bibitem[{{Roberts}(2004)}]{Roberts2004}
{Roberts}, B. 2004, in ESA Special Pub. (Paris: ESA), Vol. 547, SOHO 13 Waves,
  Oscillations and Small-Scale Transients Events in the Solar Atmosphere: Joint
  View from SOHO and TRACE, ed. {{Lacoste}, H.}, 1

\bibitem[{{Roberts}(2006)}]{Roberts2006}
{Roberts}, B. 2006, Philos. Trans. R. Soc. London A, 364, 447

\bibitem[{{Roberts} \& {Ulmschneider}(1997)}]{Roberts1997}
{Roberts}, B. \& {Ulmschneider}, P. 1997, in Lecture Notes in Phys. (Berlin:
  Springer-Verlag), Vol. 489, European Meeting on Solar Physics, ed.
  {{Simnett}, G.~M. and {Alissandrakis}, C.~E. and {Vlahos}, L.}, 75

\bibitem[{{Rosenthal} {et~al.}(2002){Rosenthal}, {Bogdan}, {Carlsson}, {Dorch},
  {Hansteen}, {McIntosh}, {McMurry}, {Nordlund}, \& {Stein}}]{Rosenthal2002}
{Rosenthal}, C.~S., {Bogdan}, T.~J., {Carlsson}, M., {et~al.} 2002, \apj, 564,
  508

\bibitem[{{Rutten} {et~al.}(2004){Rutten}, {de Wijn}, \&
  {S{\"u}tterlin}}]{Rutten2004}
{Rutten}, R.~J., {de Wijn}, A.~G., \& {S{\"u}tterlin}, P. 2004, \aap, 416, 333

\bibitem[{{Rutten} \& {Uitenbroek}(1991)}]{Rutten1991}
{Rutten}, R.~J. \& {Uitenbroek}, H. 1991, \solphys, 134, 15

\bibitem[{{Schunker} \& {Cally}(2006)}]{Schunker2006}
{Schunker}, H. \& {Cally}, P.~S. 2006, \mnras, 372, 551

\bibitem[{{Skumanich} {et~al.}(1975){Skumanich}, {Smythe}, \&
  {Frazier}}]{Skumanich1975}
{Skumanich}, A., {Smythe}, C., \& {Frazier}, E.~N. 1975, \apj, 200, 747

\bibitem[{{Solanki}(1993)}]{Solanki1993}
{Solanki}, S.~K. 1993, \ssr, 63, 1

\bibitem[{{Solanki} {et~al.}(2010){Solanki}, {Barthol}, {Danilovic}, {Feller},
  {Gandorfer}, {Hirzberger}, {Riethm{\"u}ller}, {Sch{\"u}ssler}, {Bonet},
  {Mart{\'{\i}}nez Pillet}, {del Toro Iniesta}, {Domingo}, {Palacios},
  {Kn{\"o}lker}, {Bello Gonz{\'a}lez}, {Berkefeld}, {Franz}, {Schmidt}, \&
  {Title}}]{Solanki2010}
{Solanki}, S.~K., {Barthol}, P., {Danilovic}, S., {et~al.} 2010, \apjl, 723,
  L127

\bibitem[{{Solanki} {et~al.}(2017){Solanki}, {Riethm{\"u}ller}, {Barthol},
  {Danilovic}, {Deutsch}, {Doerr}, {Feller}, {Gandorfer}, \& {Sunrise
  team}}]{Solanki2017}
{Solanki}, S.~K., {Riethm{\"u}ller}, T.~L., {Barthol}, P., {et~al.} 2017,
  \apjs, 229, 2

\bibitem[{{Solanki} {et~al.}(1991){Solanki}, {Steiner}, \&
  {Uitenbroek}}]{Solanki1991}
{Solanki}, S.~K., {Steiner}, O., \& {Uitenbroek}, H. 1991, \aap, 250, 220

\bibitem[{{Souffrin}(1972)}]{Souffrin1972}
{Souffrin}, P. 1972, \aap, 17, 458

\bibitem[{{Spruit}(1981)}]{Spruit1981a}
{Spruit}, H.~C. 1981, \aap, 98, 155

\bibitem[{{Spruit}(1982)}]{Spruit1982}
{Spruit}, H.~C. 1982, \solphys, 75, 3

\bibitem[{{Stangalini} {et~al.}(2011){Stangalini}, {Del Moro}, {Berrilli}, \&
  {Jefferies}}]{Stangalini2011}
{Stangalini}, M., {Del Moro}, D., {Berrilli}, F., \& {Jefferies}, S.~M. 2011,
  \aap, 534, A65

\bibitem[{{Stangalini} {et~al.}(2015){Stangalini}, {Giannattasio}, \&
  {Jafarzadeh}}]{Stangalini2015}
{Stangalini}, M., {Giannattasio}, F., \& {Jafarzadeh}, S. 2015, \aap, 577, A17

\bibitem[{{Stangalini} {et~al.}(2013){Stangalini}, {Solanki}, {Cameron}, \&
  {Mart{\'{\i}}nez Pillet}}]{Stangalini2013}
{Stangalini}, M., {Solanki}, S.~K., {Cameron}, R., \& {Mart{\'{\i}}nez Pillet},
  V. 2013, \aap, 554, A115

\bibitem[{{Straus} {et~al.}(2008){Straus}, {Fleck}, {Jefferies}, {Cauzzi},
  {McIntosh}, {Reardon}, {Severino}, \& {Steffen}}]{Straus2008}
{Straus}, T., {Fleck}, B., {Jefferies}, S.~M., {et~al.} 2008, \apjl, 681, L125

\bibitem[{{Suematsu}(1990)}]{Suematsu1990}
{Suematsu}, Y. 1990, in Lecture Notes in Physics, Berlin Springer Verlag, Vol.
  367, Progress of Seismology of the Sun and Stars, ed. Y.~{Osaki} \&
  H.~{Shibahashi}, 211

\bibitem[{{Taroyan} \& {Erd{\'e}lyi}(2009)}]{Taroyan2009}
{Taroyan}, Y. \& {Erd{\'e}lyi}, R. 2009, \ssr, 149, 229

\bibitem[{{Torrence} \& {Compo}(1998)}]{Torrence1998}
{Torrence}, C. \& {Compo}, G.~P. 1998, BAMS, 79, 61

\bibitem[{{Uitenbroek}(1989)}]{Uitenbroek1989}
{Uitenbroek}, H. 1989, \aap, 213, 360

\bibitem[{{Uitenbroek}(2001)}]{Uitenbroek2001}
{Uitenbroek}, H. 2001, \apj, 557, 389

\bibitem[{{Ulmschneider} {et~al.}(1991){Ulmschneider}, {Zaehringer}, \&
  {Musielak}}]{Ulmschneider1991}
{Ulmschneider}, P., {Zaehringer}, K., \& {Musielak}, Z.~E. 1991, \aap, 241, 625

\bibitem[{{Vigeesh} {et~al.}(2012){Vigeesh}, {Fedun}, {Hasan}, \&
  {Erd{\'e}lyi}}]{Vigeesh2012}
{Vigeesh}, G., {Fedun}, V., {Hasan}, S.~S., \& {Erd{\'e}lyi}, R. 2012, \apj,
  755, 18

\bibitem[{{Vigeesh} {et~al.}(2009){Vigeesh}, {Hasan}, \&
  {Steiner}}]{Vigeesh2009}
{Vigeesh}, G., {Hasan}, S.~S., \& {Steiner}, O. 2009, \aap, 508, 951

\bibitem[{{Wang} {et~al.}(2003){Wang}, {Solanki}, {Curdt}, {Innes}, {Dammasch},
  \& {Kliem}}]{Wang2003}
{Wang}, T.~J., {Solanki}, S.~K., {Curdt}, W., {et~al.} 2003, \aap, 406, 1105

\bibitem[{{Wedemeyer-B{\"o}hm} {et~al.}(2007){Wedemeyer-B{\"o}hm}, {Steiner},
  {Bruls}, \& {Rammacher}}]{Wedemeyer-Bohm2007}
{Wedemeyer-B{\"o}hm}, S., {Steiner}, O., {Bruls}, J., \& {Rammacher}, W. 2007,
  in ASP Conf. Ser., Vol. 368, The Physics of Chromospheric Plasmas, ed.
  P.~{Heinzel}, I.~{Dorotovi{\v c}}, \& R.~J. {Rutten} (San Francisco, CA:
  ASP), 93

\end{thebibliography}

\end{document}